\documentclass[10journal,compsoc]{IEEEtran}
\usepackage{amsmath,amsfonts}
\usepackage{algorithmic}
\usepackage{algorithm}
\usepackage{array}
\usepackage[caption=false,font=normalsize,labelfont=sf,textfont=sf]{subfig}
\usepackage{textcomp}
\usepackage{stfloats}
\usepackage{url}
\usepackage{verbatim}
\usepackage{graphicx}
\usepackage{cite}
\usepackage{inconsolata}
\usepackage{textcomp}
\usepackage{booktabs}
\usepackage[hidelinks]{hyperref}

\begin{document}

\title{Token Interdependency Parsing (Tipping) - A Fast and Accurate Statistical Log Parser}

%\title{Tipping - Parallel Token Interdependency Analysis for Fast and Accurate Log Parsing}

%\title{Token Interdependency Log Parsing (Tipping)}

\author{Shayan Hashemi, M3S, University of Oulu, Finland \\ Mika Mäntylä, Department of Computer Science, University of Helsinki, Finland \& M3S, Oulu, Finland}
        % <-this % stops a space

\markboth{Journal of \LaTeX\ Class Files,~Vol.~14, No.~8, August~2021}%
{Shell \MakeLowercase{\textit{et al.}}: A Sample Article Using IEEEtran.cls for IEEE Journals}

\IEEEpubid{0000--0000/00\$00.00~\copyright~2021 IEEE}
% Remember, if you use this, you must call \IEEEpubidadjcol in the second

\maketitle

\begin{abstract}
%Motivation
In the last decade, an impressive increase in software adaptations has led to a surge in log data production, making manual log analysis impractical and establishing the necessity for automated methods. Conversely, most automated analysis tools include a component designed to separate log templates from their parameters, commonly referred to as a "log parser".

%Objective
This paper aims to introduce a new fast and accurate log parser, named "Tipping".
%Method
Tipping combines rule-based tokenizers, interdependency token graphs, strongly connected components, and various techniques to ensure rapid, scalable, and precise log parsing. Furthermore, Tipping is parallelized and capable of running on multiple processing cores with close to linear efficiency. We evaluated Tipping against other state-of-the-art log parsers in terms of accuracy, performance, and the downstream task of anomaly detection.

%Results
Accordingly, we found that Tipping outperformed existing methods in accuracy and performance in our evaluations. More in-depth, Tipping can parse 11 million lines of logs in less than 20 seconds on a laptop machine. Furthermore, we re-implemented a parallelized version of the past IpLom algorithm to demonstrate the effect of parallel processing, and it became the second-fastest parser.

%Conclusion
As logs keep growing in volume and complexity, the software engineering community needs to ensure automated log analysis tools keep up with the demand, being capable of efficiently handling massive volumes of logs with high accuracy. Tipping's robustness, versatility, efficiency, and scalability make it a viable tool for the modern automated log analysis task.

 %The vast amount and complexity of these logs make manual analysis increasingly difficult, 

%These tools utilize advanced algorithms, including machine learning and artificial intelligence, to handle large-scale log data more efficiently and accurately, improving system management and security.
%Objective

%Method
%Results
%Conclusion
%The log parser, a key component in the analysis pipeline, converts raw log data into a structured format for deeper analysis. 
%Method
%four benchmarks (LogPai, LogPM, and two LogLead benchmarks), demonstrating its capability to handle extensive log data swiftly and accurately, making it a valuable tool in modern digital infrastructure.

%1.6s for one million lines

\end{abstract}


\section{Introduction}
% 1) RQ1 - What is the mathematical background of Tipping
% 2) RQ2 - How does Tipping perform in terms of parsing accuracy?
% 3) RQ3 - How does Tipping perform in terms of parsing speed?
% - Here. We could do both VMs with 28 CPUs and laptop 4cpus to see how the
% parallelization of tipping helps.
% 4) RQ4 - How does Tipping perform in terms of the downstream task of
% anomaly detection?
\IEEEPARstart{I}{n}
recent years, breakthroughs in software development have led to an exponential increase in log generation, as complex systems and applications continuously produce vast amounts of log data. These logs provide insights into system performance, user behaviours, and potential security threats. However, the sheer volume and speed at which this data is produced have made it increasingly challenging for human analysts to manually process and interpret these logs effectively. As software systems grow more sophisticated and interconnected, the deluge of data, including transaction details, system statuses, and operational anomalies, continues to expand.

%This surge in data has made manual analysis impractical due to time constraints and the potential for human error in handling such extensive datasets, necessitating the adoption of automated log analysis tools. Automated tools employ advanced algorithms, including those powered by machine learning and artificial intelligence, to efficiently parse, analyze, and model log data on a large scale. These technologies not only speed up the analysis process but also enhance its accuracy and effectiveness in some cases, enabling organizations to swiftly identify and address issues, optimize system performance, and strengthen security measures. As a result, automated log analysis has become an indispensable tool in modern digital infrastructure management.

The log parser is one of the most crucial components in any log analysis pipeline. This tool is the foundational mechanism that interprets raw log messages into a structured format, making it crucial for further analysis. Log parsing, sometimes referred to as log clustering, is a method aimed at simplifying the complexity of free-form log messages. This process involves differentiating between a log message's constant and variable parts; the constant part (template) represents the event description, while the variable part (parameters) includes dynamic data specific to each event. As an illustration, in the log message "Socket connected successfully to 10.10.1.10", the phrase "Socket connected successfully to" is the template, which indicates an event. At the same time, "10.10.1.10" represents the replaceable parameter that may vary with each specific event occurrence.

Furthermore, log parsing plays a vital role in enhancing software management and security across various fields\cite{logpm}. It is essential for \textbf{anomaly detection} \cite{sialog,loganomaly,deeplog,cnnlog,monika}, aiding in the identification and prediction of unusual behaviors through advanced algorithms. For \textbf{debugging} \cite{logdebugging1,logdebugging2}, it enables developers to trace errors back to specific lines of code or external factors. Log parsing is used to identify bottlenecks, improving application efficiency and speed in \textbf{performance analysis} \cite{netlogger,loggingperformance}. \textbf{Security analysis} \cite{securitylog1,securitylog2} leverages log parsing to detect potential threats by analyzing patterns of suspicious activity. \textbf{Real-time software monitoring} \cite{logsbusiness} utilizes parsed logs to gain insights into user behaviour and system performance, facilitating issue detection. \textbf{Root cause analysis} \cite{rcalog1,rcalog2} uses log parsing to model common event patterns, helping to identify the origins of software issues. Log parsing supports \textbf{audit trails} \cite{audit} by tracking data access and modifications, ensuring compliance and security. Each of these applications underscores the versatility and indispensability of log parsing in modern software engineering. The pivotal importance of log parsing has prompted the creation of numerous tools aimed at optimizing this process \cite{brain,drain,spell,molfi,logram,ulp,loghub,loghubtools}

In log parsing, performance and accuracy stand as the twin pillars essential to its effectiveness. Performance refers to the speed at which log data can be processed and analyzed, a critical factor when dealing with large volumes of logs. Accuracy, however, involves the precision with which the log parser can identify and separate the template and parameter parts from each other, which is crucial for meaningful interpretation and making the most of downstream components. However, these two often find themselves at odds. Enhancing accuracy typically involves more complex processing techniques that are computationally intensive and slower. This inverse relationship poses a significant challenge: striving for higher accuracy can compromise the speed.

Parallel processing refers to distributing tasks across multiple processors or cores to be executed simultaneously, enhancing speed as a result. Accordingly, log parsing can be improved by leveraging parallel processing, as parallel processing allows for handling extensive datasets at a faster rate while allowing for the incorporation of complex parsing algorithms that ensure high accuracy. However, implementing parallel processing comes with its own set of challenges, such as complexity, mutual exclusion, and ensuring data consistency and synchronization.

We offer a novel parsing algorithm named "Tipping," which employs a rule-based tokenizer, an interdependency token graph, strongly connected components, and the map-reduce parallelism paradigm to optimize performance and efficiently parse logs. Such a structure allows for rapid execution times that scale effectively with the number of CPU cores utilized. Additionally, Tipping is designed with various options, providing extensive flexibility for fine-tuning according to specific needs. This ensures that our algorithm is adaptable to a wide range of environments and requirements.

Finally, we demonstrate that the Tipping algorithm outperforms state-of-the-art methods across four benchmarks, highlighting its accuracy and performance. Tipping yields the fastest parsing times in the LogHub2k, LogHub2.0, and LogPM benchmarks while maintaining accuracy, and processes a massive volume of logs faster than other methods in the LogLead for computational efficiency. Additionally, Tipping excels in LogLead's anomaly detection downstream task as well. These results underscore Tipping's robustness and adaptability, making it a performant, accurate, and versatile tool for various log analysis applications.

\section{Related works}

In a brilliant study, Drain \cite{drain} is designed to process logs in a streaming and timely manner. Unlike traditional log parsers focusing on offline batch processing and struggling with the rapidly increasing volume of logs, Drain introduced a fixed-depth parse tree to accelerate the parsing process. This parse tree, crucial for guiding the log group search process, encodes specially designed parsing rules within its nodes. Its fixed-depth tree structure and specially designed parsing rules significantly enhance the speed and accuracy of log parsing, making it an excellent tool for online log parsing.

Another study introduces the Brain\cite{brain}, an innovative log parsing approach that leverages a bidirectional parallel tree to transform semi-structured logs into a structured format, aimed at overcoming the limitations of existing parsers in software systems. By identifying the longest common pattern across logs as a likely component of the log template, Brain efficiently groups logs and hierarchically integrates constant words into these patterns to form complete templates. This method significantly enhances parsing accuracy and efficiency across various software logs. Furthermore, Brain's process encompasses preprocessing for word splitting and variable filtering, initial group creation based on common patterns, and using a bidirectional tree to systematically add constant words, culminating in the generation of structured log templates. Brain offers improved stability and processing speed for large volumes of log data.

NuLog, introduced in \cite{nulog}, is a self-supervised log parsing technique leveraging transformer architecture to address the challenge of parsing massive volumes of semi-structured logs generated by large-scale software systems. Unlike traditional methods that rely on heuristics or manual rule extraction, NuLog uses masked language modeling (MLM) to predict the presence of words in log messages based on their context, distinguishing between constant and variable parts without requiring domain knowledge. The methodology encompasses tokenization, masking, and a multi-head self-attention mechanism with positional encoding and a feed-forward network, outputting log templates alongside numerical vector summarizations. This enables high parsing accuracy and supports downstream anomaly detection tasks in both supervised and unsupervised scenarios. NuLog's approach significantly advances automated log analysis, offering broad applicability and minimal human intervention.

Another ingenious paper presents Spell\cite{spell}, a novel streaming log parser that leverages the Longest Common Subsequence (LCS) method for parsing system event logs in real-time into structured data. Moreover, Spell dynamically extracts log patterns and updates its log structure by comparing incoming log entries against stored LCS sequences within an LCSMap data structure. Furthermore, It utilizes a pre-filtering step using a prefix tree and a simple loop approach to efficiently identify existing message types, significantly reducing the computational load by applying LCS computation selectively. Spell is considered an effective solution for online log monitoring and analysis, offering immediate insights into system behavior and enhancing log management capabilities.

MoLFI, introduced in \cite{molfi}, proposes an innovative approach by framing it as a multi-objective optimization problem, efficiently solved using the NSGA-II \cite{nsgaii} algorithm to balance the frequency and specificity of templates. This method requires no domain knowledge to address the challenges of rapidly evolving log formats. MoLFI produces a Pareto optimal set of templates that adeptly navigates the trade-offs between template commonality and uniqueness through multiple steps of preprocessing, solution encoding, and iterative refinement using genetic operations. Overall, MoLFI introduced a completely different approach compared to other well-known log parsers at the time.

Logram\cite{logram} is another log parsing method that employs n-gram dictionaries to address the log parsing problem efficiently. By leveraging the intuition that frequently occurring n-grams likely represent static text and rare n-grams indicate dynamic variables, Logram generates n-gram dictionaries from log messages and uses these dictionaries to parse and distinguish between static and dynamic components of logs. This process involves preprocessing logs to extract tokens, creating dictionaries for efficient queries, and employing an automated method to determine the occurrence threshold for dynamic variable identification. In their own evaluations, Logram showed improved parsing accuracy and efficiency.

ULP (Unified Log Parser)\cite{ulp} is a log parsing tool that enhances accuracy and efficiency through a novel combination of string matching and local frequency analysis. ULP effectively differentiates between static and dynamic components of log messages by grouping log events based on string similarity and then applying frequency analysis within these groups. This method outperformed its rival competition at the time. The parser involves preprocessing to identify trivial dynamic tokens, grouping similar events for targeted analysis, and generating templates through local frequency insights, showcasing a significant improvement in parsing large log files and offering a solution for complex log analysis tasks.

The log parser LenMa (Length Matters) \cite{lenma} introduces a unique clustering approach, leveraging the length of words within each message to distinguish between fixed and variable components, thereby facilitating efficient online template generation. LenMa simplifies the process by employing word length vectors to compute similarity scores between new log entries and existing clusters using cosine similarity. This innovative approach enables real-time, dynamic clustering of log messages, efficiently adapting to the continuous influx of logs. LenMa proves to be an effective and straightforward method for managing vast volumes of log data in a scalable solution.

IpLom (Iterative Partitioning Log Mining) \cite{makanju2009clustering,makanju2011lightweight} is a multi-step log clustering approach that hierarchically partitions log messages into different groups. The first step of partitioning divides log messages into groups based on the number of tokens. The second step further processes each of these groups by splitting them based on the token position with the least variability, i.e., they are the most likely to be templates. The third step searches for a bijective relationship with token positions that co-vary. Although presented already 15 years ago, IpLom has shown promising accuracy and speed in independent tests \cite{logpm}. The main drawback of IpLom is that varying numbers of tokens in log parameters can cause incorrect groups, and the algorithm does not include the ability to correct these groups.

AEL (Abstracting Execution Logs) \cite{jiang2008automated} is based on four hierarchical steps: anonymize, tokenize, categorize, and reconcile. The anonymized step uses external heuristics to convert log message parameters to generic tokens. The tokenize step uses the number of tokens and parameters from the previous step to group log messages. The categorization step goes through existing groups and performs a token-by-token comparison and clustering. Finally, the reconcile step tries to account for the lack of heuristics of the anonymized step and merges log clusters if too many execution events have been created. AEL has also shown good accuracy and speed in independent tests \cite{logpm}. The main drawback of the algorithm is the need to have A-priori heuristics on how to recognize log message parameters.

\section{Proposed Method}
 Tipping is divided into two phases. During the initial phase (Sec. \ref{sec:tokenizer} to \ref{sec:occandcooc}), it gathers information from log messages. In the subsequent phase (Sec. \ref{sec:interdepgraph} to \ref{sec:parammask}), the collected information is employed to cluster messages, find templates, and generate parameter masks. This two-phase approach enhances the algorithm's accuracy and allows for parallel processing.

\begin{figure*}[t]
\label{fig:tip-overview}
\centering
\includegraphics[width=\linewidth]{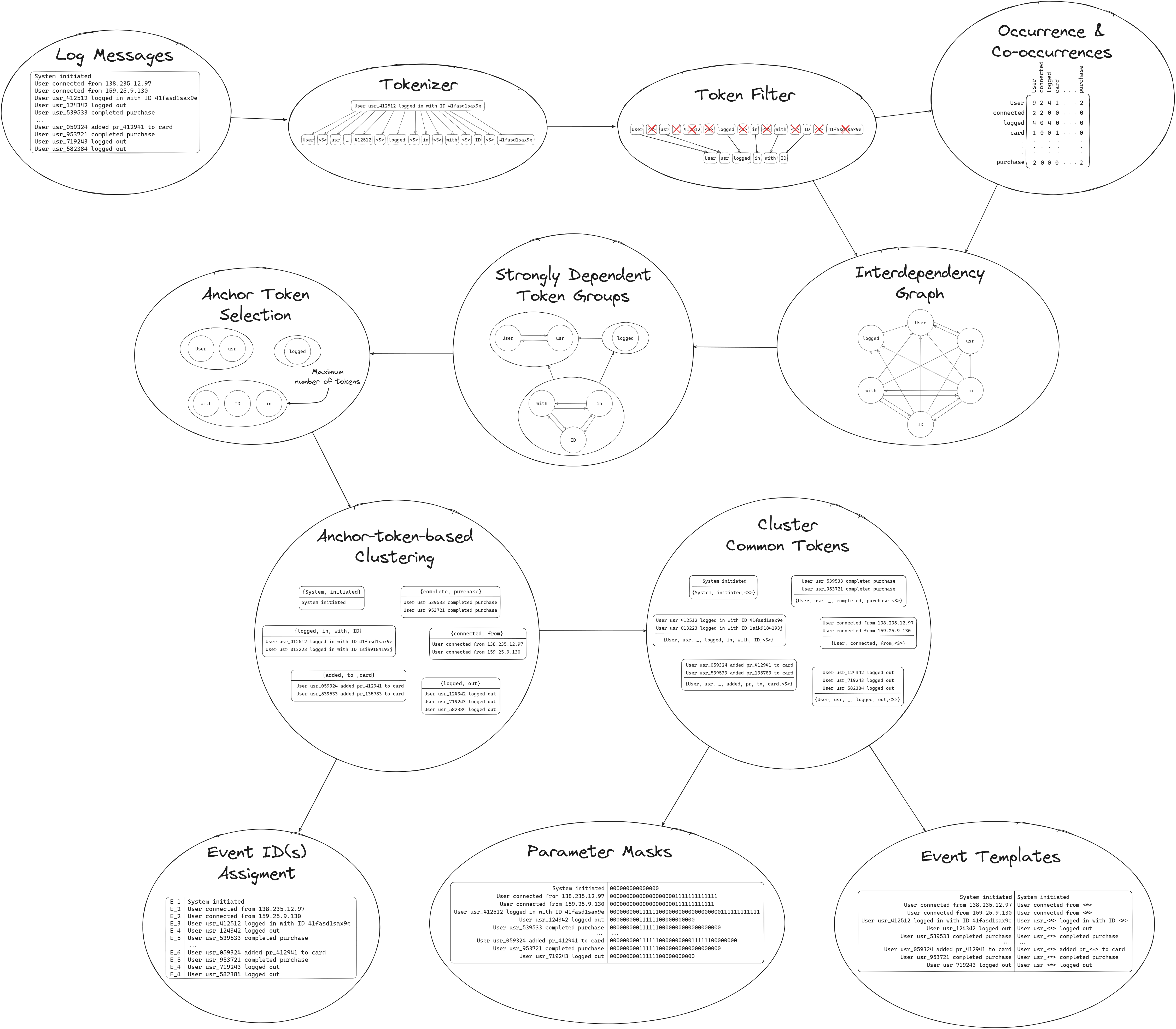}
\centering
\caption{An overall view of Tipping's components, internal workflow, relation, and execution order.}
\end{figure*}

Figure \ref{fig:tip-overview} visualize an overview of our approach, while the following subsections explain each phase in detail. 
The tokenizer, which divides log messages into smaller parts, is discussed at first. Then, interdependency graph construction and anchor token identification are described. The final steps involve clustering messages based on these anchor tokens and generating event templates and parameter masks.
\subsection{Tokenizer}
\label{sec:tokenizer}
The first component of our parser is the tokenizer, which is common in almost every NLP task. The tokenizer breaks down log messages into smaller substrings that make the data more suitable for algorithms to process. Additionally, our approach allows for domain knowledge to be passed from users to the tokenizer using regular expressions.

Accordingly, the tokenizer takes three inputs: white special patterns, black special patterns, and symbols. While the special tokens are optional, the symbol set has the default value of $\{  ( \quad ) \quad [ \quad ] \quad \{ \quad \} \quad = \quad , \quad * \}$.

The white special patterns are a list of regular expressions that should always be recognized as templates. Conversely, black special patterns should be recognized as parameters.

Special patterns allow for more nuanced control in the parsing process, resolving problems such as maintaining template-shared tokens\cite{logpm} and execution environment bias parameters\cite{logpm}. Lastly, the symbols consist of a set of punctuation characters that, alongside whitespace characters, serve as delimiters in the tokenization process.

The tokenizer initiates its process by scanning the log message to identify and mark all occurrences of white special patterns. Subsequently, the algorithm examines the remaining parts to locate and mark black special patterns. Then, the algorithm iterates over the rest of the segments and uses symbols as delimiters to split the message.

\subsection{Token Filter}
\label{sec:tokenfilter}
The token filter receives tokenized messages and filters out tokens based on user-defined rules. This not only allows for better leverage of domain knowledge but also results in fewer tokens and less processing, as the results for the next components. By default, the token filter only keeps the white special tokens and purely alphabetic tokens.

% \textbf{Parallelization}: 
% Since the tokenizer and token filter are rule-based, one message's output is completely independent of the other messages. This independence allows each message to be processed in isolation without any dependencies or interactions with others. Consequently, this architecture enables straightforward parallelization, as multiple messages can be processed simultaneously without risking inconsistencies or conflicts in the output.

% \begin{figure*}[t]
% \centering
% \includegraphics[width=\linewidth]{images/tokenizer.png}
% \caption{An example of a tokenizer}
% \end{figure*}

%Above very clear

\subsection{Occurences and Co-occurrences}
\label{sec:occandcooc}
For further next components, keeping track of both token occurrences and their co-occurrences is essential.

Token occurrence refers to the number of times that a particular token appears within a given dataset. On the other hand, co-occurrence involves tracking instances where two tokens appear together within the same message. Both occurrence and co-occurrence could be stored in a single sparse matrix data structure. However, a hash map is faster and more memory efficient, as row- or column-wise access is not required. Within the hash map data structure, keys are single or two-member token sets, indicating occurrences or co-occurrences, while the value is the count for the keys.

% \textbf{Parallelization}:
% The parallelization of occurrences and co-occurrences is straightforward as the problem could be mapped to a counting problem, which could be easily resolved in the map-reduce paradigm.

\subsection{Interdependency Graph}
\label{sec:interdepgraph}
Before defining the interdependency graph, it's essential to define dependency in this context. Token dependency, inspired by the concept of support in recommender systems\cite{lu2015recommender} and conditional probability, is defined as the ratio of log messages in which both tokens appear together to the number of log messages that contain the dependent one. In other words, the dependency of the token $A$ on $B$ is the co-occurrence(s) of $A$ and $B$ together divided by the total occurrences of $A$. To formalize this, if $S_A$ represents the set of messages containing token $A$ while $S_B$ denotes those with token $B$, the dependency of token $A$ on token $B$ can be mathematically expressed as:
$$
dep(A,B) = \frac{|S_A \cap S_B|}{|S_A|}
$$
Accordingly, the interdependency graph is a directed graph constructed for each log message individually, where nodes (vertices) are tokens and edges (arcs) are the dependence of the source nodes' tokens on destination nodes' tokens.

\subsection{Strongly Dependent Token Group}
\label{sec:tokengroup}
A strongly dependent token group is a subset of tokens within a message characterized by their direct or indirect mutual dependence on each other. Conversely, tokens outside this subset either do not depend on the tokens within the group or the group tokens do not depend on them. It is worth mentioning that there is at least one token group in a message, while the maximum number is equal to the number of unique tokens in that message.

The intuition behind dependency grouping is that when token $A$ depends on token $B$, and $A$ is identified as part of the template, inevitably, $B$ is also part of the template, as token $A$ is unlikely to occur without token $B$. Therefore, tokens within the same token group are expected to appear together.

A threshold hyper-parameter, denoted as $\theta$, is required to compute strongly dependent token groups (reasons are going to be discussed in the next paragraph). The process begins by constructing a token interdependency graph from the message. Then, edges with weights below the $\theta$ are disregarded. Then, strongly dependent token groups are identified using strongly connected component algorithms, such as the Tarjan \cite{tarjan} or Kosaraju-Sharir \cite{sharir1981strong} algorithm.

There are two reasons for disregarding edges below the threshold. First, it helps to get rid of low dependencies often caused by noise. Second, since any two tokens within a single message are mentioned together at least once, a baseline of dependency always exists among all tokens of a message, leading to a fully connected interdependency graph. Therefore, since strongly connected components algorithms rely only on the presence of edges, not their weights, it will result in only one group.

\subsection{Anchor Tokens Selection}
\label{sec:anchortok}
After identifying strongly dependent token groups, the largest one is selected as the anchor group, and its members are designated as the anchor tokens. Algorithm \ref{alg:anchor_tokens} outlines all the steps required to select the anchor token.
% \textbf{Parallelization}:
% Despite constructing the interdependency graph, selecting strongly connected components, and selecting anchor tokens occurring sequentially in a pipeline, the results of different pipelines are independent of each other. This independence allows for multiple pipeline executions in parallel.

% \begin{figure}[]
% \centering
% \includegraphics[width=\linewidth]{images/connected components.pdf}
% \caption{An example of a tokenizer}
% \end{figure}

\begin{algorithm}[H]
\caption{Anchor token identification given message $m$ and threshold $\theta$.}
\label{alg:anchor_tokens}
\begin{algorithmic}
\STATE 
\STATE {$t \gets tokenize(m)$}
\STATE {$t^{'} \gets filter(t)$}
\STATE {$g \gets interdep\_graph(t^{'})$}
\FOR {$e \in edges(g)$}
    \IF {$e < \theta$}
        \STATE {$remove(e)$}
    \ENDIF
\ENDFOR
\STATE {$anchor \gets \{\emptyset\}$}
\FOR {$c \in connected\_components(g)$}
    \IF {$|anchor| < |c|$}
        \STATE {$anchor \gets c$}
    \ENDIF
\ENDFOR
\RETURN {$anchor$}
\end{algorithmic}
\end{algorithm}

\subsection{Anchor-token-based Clustering}
\label{sec:clustering}
After identifying anchor tokens, messages are clustered into distinct groups based on their anchor tokens. Then, event IDs could be obtained from clusters by computing the cluster hash or its index. Should the desired outcome be solely the event IDs, the algorithm concludes at this juncture. However, if there is a request for templates or parameter masks in addition to event IDs, the algorithm proceeds to produce the next three stages.

\subsection{Cluster Common Tokens}
\label{sec:commontok}

Common tokens within each cluster are those present across all messages in the cluster, effectively representing the intersection of tokens in all messages. Subsequently, these common tokens are later leveraged to generate event templates and parameter masks.

\subsection{Event Templates}
\label{sec:eventtemplates}

\begin{table*}[!t]
\caption{An example of a situation where a single event may have different template representations.}
\label{tab:proxexam}
\centering
\begin{tabular}{rl}
\toprule
Message 1 &\texttt{\textbf{chrome.exe *64} - \textbf{c.cnzz.com:80} close, \textbf{0} bytes sent, \textbf{0} bytes received, lifetime \textbf{00:10}} \\
Template 1 & \texttt{<*> close, <*> bytes sent, <*> bytes received, lifetime <*>}\\
\midrule
Message 2 &\texttt{\textbf{chrome.exe *64} - \textbf{nos.netease.com:443} close, \textbf{329} bytes sent, \textbf{3490} bytes \textbf{(3.40 KB)} received, lifetime \textbf{01:00}} \\
Template 2 & \texttt{<*> close, <*> bytes sent, <*> bytes <*> received, lifetime <*>}\\
\midrule
Message 3 &\texttt{\textbf{chrome.exe *64} - \textbf{css.sohu.com:80} close, \textbf{1463} bytes \textbf{(1.42 KB)} sent, \textbf{766} bytes received, lifetime \textbf{00:22}} \\
Template 3 & \texttt{<*> close, <*> bytes <*> sent, <*> bytes received, lifetime <*>}\\
\midrule
Message 4 &\texttt{\textbf{chrome.exe *64} - \textbf{manhua.163.com:443} close, \textbf{8544} bytes \textbf{(8.34 KB)} sent, \textbf{102059} bytes \textbf{(99.6 KB)} received, lifetime \textbf{00:08}} \\
Template 4 & \texttt{<*> close, <*> bytes <*> sent, <*> bytes <*> received, lifetime <*>}\\
\bottomrule
\end{tabular}
\end{table*}

Tipping produces a set of templates for each event to handle different template variations of the event. For instance, in the example provided by Table \ref{tab:proxexam}, all messages are representations of the same event. However, due to optional parameters within the log message structure\cite{logpm}, the templates are different from each other.

The process starts with a cluster of log messages and their common tokens. Initially, it involves processing these messages to create preliminary templates, where common tokens are kept and others are replaced with placeholders. Then, consecutive placeholders are merged into a single one. To make the template representable as a string, placeholders should be replaced by a parameter symbol, such as ``\textless*\textgreater''. The final step aggregates these templates into a set, producing a set of templates for that cluster.

\subsection{Parameter Masks}
\label{sec:parammask}
After identifying the common tokens, we initialize the parameter mask with an empty string for each message. Then, the parameter mask is populated by iterating over the message tokens, appending $n$ zeros if the token is present in the common tokens or $n$ ones otherwise, where $n$ is the number of characters in each token. Repeating this for all messages generates parameter masks for the entire cluster.

\section{Experiments}
Given that our algorithm relies on parallel computing, we implemented our algorithm in Rust due to its fearless concurrency, exceptional performance, and seamless interoperability with other languages. However, since Python is the dominant language in the data science and AI space, we also released a Python binding for our package (the core is still written in Rust), which will be utilized in the upcoming experiment. The Tipping Python package can be installed from \textit{pypi}\footnote{\url{https://pypi.org/project/tipping}}, while the Rust package is available on \textit{crates.io}\footnote{\url{https://crates.io/crates/tipping-rs}}.

Unless explicitly mentioned, all experiments are performed on a 64-bit OpenStack virtual machine featuring 16 Intel Broadwell 2 GHz virtual CPUs capable of boosting to high frequencies and 234 GiB of ECC memory, optimized for parallel processing and data integrity. All virtual disks are formatted utilizing EXT4 and XFS file systems for efficient data handling. Furthermore, the VM is equipped with two NVIDIA Tesla P100 GPUs (16 GB each). It is worth mentioning that none of the experiment methods utilizes a GPU.

We performed four experiments, see Sections \ref{sec:logpai} to \ref{sec:loglead2}, to answer the following research questions:

\begin{itemize}
    \item \textbf{RQ1}: How accurately and efficiently does Tipping perform in LogHub2k and LogHub2.0 benchmarks? See Section \ref{sec:logpai}.
    \item \textbf{RQ2}: How accurately and efficiently does Tipping perform in the LogPM benchmark? See Section \ref{sec:logpm}.
    \item \textbf{RQ3}: How sensitive is Tipping to its hyperparameters, such as $\theta$ and special patterns? See Section \ref{sec:hypersens} 
    \item \textbf{RQ4}: How fast is Tipping according to the LogLead benchmark? See Section \ref{sec:loglead1} 
    \item \textbf{RQ5}: What is the impact of Tipping as a parser on the downstream task? Experiment in Section \ref{sec:loglead2}
\end{itemize}

Finally, since our experiments yield multiple tables as results, yet they are presented in an aggregated format due to better readability and understandability, we have made the non-aggregated results available in a dedicated GitHub repository\footnote{\url{https://github.com/M3SOulu/tipping-detailed-result-tables}} for further transparency.

\subsection{LogHub Benchmark}
\label{sec:logpai}

\subsubsection{Experiment depiction}

LogPai's benchmark \cite{loghubtools}, published in 2019 and followed up by another study \cite{loghub2} in 2024, has become a canonical accuracy evaluation benchmark for log parsers since its introduction. The first benchmark (LogHub2K) comprises 2,000 manually labeled samples across 16 public open datasets. It primarily uses Group Accuracy (GA) and the $F_1$ score as metrics. Additionally, it includes parsing time as a performance metric. However, it is observed that performance might not be fully representative of real-world scenarios in the first benchmark due to the limited number of samples in each dataset \cite{logpm}. On the other hand, during the follow-up study \cite{loghub2}, the second benchmark (LogHub2.0) extended the datasets' coverage. Furthermore, they added new metrics such as $F_1$ score of Group Accuracy (FGA) and Template Accuracy (TA), which contains Precision Template Accuracy (PTA), Recall Template Accuracy (RTA), and $F_1$ Template Accuracy (FTA).

In addition to the benchmark metrics, we introduced two new ones. The first metric,  Parsed Dataset (PD), unveils the number of datasets that were successfully parsed without crashing or running out of memory within the time window defined by \cite{loghub2} (12 hours).

The second metric, Normalized Parsing Time (NPT), reveals the total time spent on the benchmark (parsing all datasets), normalized by Tipping's total time to demonstrate the parsing speed compared to Tipping.

In favor of transparency, we made the implementation, which is a fork of LogHub2.0, publicly available\footnote{\url{https://github.com/M3SOulu/loghub-2.0}}. Furthermore, we intend to merge this with the original repository upon acceptance.

\subsubsection{Experiment results}

In Table \ref{tab:loghub2k}, Tipping outperforms other methods across multiple key metrics. It achieves the highest Group Accuracy (GA) of 0.97 and $F_1$ score of Group Accuracy (FGA) of 0.93, far exceeding the averages of 0.71 and 0.58, respectively. Similarly, Tipping leads in Parsing Accuracy (PA) at 0.31, as well as in Precision Template Accuracy (PTA), Recall Token Accuracy (RTA), and $F_1$ score Template Accuracy (FTA).

\begin{table*}
\centering
\caption{The results of the LogHub 2k benchmark. The following acronyms are used: GA (Group Accuracy), FGA ($F_1$ score of Group Accuracy), PA (Parsing Accuracy), PTA (Precision Template Accuracy), RTA (Recall Template Accuracy), FTA ($F_1$ score of Template Accuracy), PD (Parsed Datasets), and NTP (Normalized Parsing Time).}
\label{tab:loghub2k}
\begin{tabular}{lrrrrrrrr}
\toprule
Method & GA & FGA & PA & PTA & RTA & FTA & PD & NPT \\
\midrule
AEL & 0.81 & 0.68 & 0.28 & 0.22 & 0.26 & 0.23 & 14 & 1.19 \\
Drain & 0.85 & 0.74 & 0.31 & 0.28 & 0.29 & 0.28 & 14 & 3.78 \\
IPLoM & 0.77 & 0.77 & 0.20 & 0.17 & 0.16 & 0.16 & 14 & 2.75 \\
LFA & 0.65 & 0.65 & 0.22 & 0.14 & 0.14 & 0.13 & 14 & 2.53 \\
LenMa & 0.77 & 0.58 & 0.13 & 0.11 & 0.18 & 0.13 & 14 & 44.54 \\
LogCluster & 0.64 & 0.30 & 0.12 & 0.07 & 0.18 & 0.09 & 14 & 2.43 \\
LogMine & 0.73 & 0.45 & 0.21 & 0.13 & 0.18 & 0.13 & 14 & 25.77 \\
LogSig & 0.50 & 0.34 & 0.11 & 0.05 & 0.04 & 0.04 & 14 & 135.52 \\
Logram & 0.53 & 0.41 & 0.17 & 0.12 & 0.14 & 0.12 & 14 & 1.55 \\
MoLFI & 0.62 & 0.59 & 0.10 & 0.09 & 0.11 & 0.09 & 14 & 205.58 \\
SHISO & 0.68 & 0.65 & 0.09 & 0.13 & 0.13 & 0.13 & 14 & 19.12 \\
SLCT & 0.61 & 0.36 & 0.26 & 0.18 & 0.12 & 0.13 & 14 & 9.93 \\
Spell & 0.77 & 0.60 & 0.22 & 0.16 & 0.19 & 0.17 & 14 & 4.64 \\
Tipping & 0.97 & 0.93 & 0.31 & 0.36 & 0.35 & 0.35 & 14 & 1.00 \\
\midrule
Average & 0.71 & 0.58 & 0.20 & 0.16 & 0.18 & 0.16 & 14 & 32.88 \\
\bottomrule
\end{tabular}
\end{table*}

Furthermore, Tipping stands out in computational efficiency, achieving the fastest parsing time while other methods, on average, take 30 times more in the LogHub2k benchmark. This efficiency gap is even more widened when compared to MoLFI (205 times more) and LogSig (135 times more).

In Table \ref{tab:loghub2.0}, Tipping continues to demonstrate strong performance, maintaining competitive metrics across the board in the LogHub2.0 benchmark. While its GA of 0.79 is lower than the top performer, AEL (0.86) performs better than the average (0.62). Its FGA is the highest at 0.74, being closely pursued by Drain (0.73), scoring significantly better than the 0.46 average. Furthermore, while not performing as the best, Tipping achieves a respectable over-average PA of 0.33. On the contrary, Tipping consistently shows top performance in the template accuracy metrics, including PTA (0.35), RTA (0.38), and FTA (0.36).

\begin{table*}
\centering
\caption{The results of the LogHub 2.0 benchmark. The following acronyms are used: GA (Group Accuracy), FGA ($F_1$ score of Group Accuracy), PA (Parsing Accuracy), PTA (Precision Template Accuracy), RTA (Recall Template Accuracy), FTA ($F_1$ score of Template Accuracy), PD (Parsed Datasets), and NPT (Normalized Parsing Time).}
\label{tab:loghub2.0}
\begin{tabular}{lrrrrrrrr}
\toprule
Method & GA & FGA & PA & PTA & RTA & FTA & PD & NPT \\
\midrule
AEL & 0.86 & 0.56 & 0.44 & 0.24 & 0.36 & 0.25 & 13 & 39.09 \\
Drain & 0.84 & 0.55 & 0.47 & 0.26 & 0.40 & 0.28 & 14 & 5.43 \\
IPLoM & 0.79 & 0.61 & 0.19 & 0.12 & 0.15 & 0.12 & 14 & 4.10 \\
LFA & 0.60 & 0.52 & 0.25 & 0.10 & 0.12 & 0.10 & 14 & 3.88 \\
LenMa & 0.82 & 0.47 & 0.24 & 0.14 & 0.26 & 0.16 & 10 & 120.13 \\
LogCluster & 0.57 & 0.04 & 0.11 & 0.01 & 0.16 & 0.01 & 14 & 2.05 \\
LogMine & 0.73 & 0.33 & 0.25 & 0.10 & 0.24 & 0.11 & 8 & 203.64 \\
LogSig & 0.18 & 0.16 & 0.09 & 0.01 & 0.01 & 0.01 & 11 & 98.71 \\
Logram & 0.34 & 0.20 & 0.20 & 0.04 & 0.24 & 0.05 & 10 & 136.49 \\
MoLFI & 0.59 & 0.35 & 0.11 & 0.03 & 0.09 & 0.04 & 11 & 106.59 \\
SHISO & 0.51 & 0.47 & 0.13 & 0.14 & 0.17 & 0.15 & 13 & 65.93 \\
SLCT & 0.40 & 0.27 & 0.24 & 0.10 & 0.23 & 0.11 & 11 & 87.88 \\
Spell & 0.70 & 0.33 & 0.22 & 0.07 & 0.16 & 0.08 & 12 & 102.86 \\
Tipping & 0.79 & 0.74 & 0.33 & 0.34 & 0.36 & 0.34 & 14 & 1.00 \\
\midrule
Average & 0.62 & 0.40 & 0.23 & 0.12 & 0.21 & 0.13 & 12.07 & 69.84 \\
\bottomrule
\end{tabular}
\end{table*}

In terms of parsing speed, Tipping shows the fastest performance. While being 69 times faster than the average, it is twice as fast as the second fastest (LogCluster). Furthermore, Tipping is among the five parsers that are capable of parsing all 14 datasets in the LogHub2.0 benchmark.

Overall, Tipping demonstrates promising results in both LogHub2k and LogHub2.0 benchmarks, producing comparatively accurate outputs with efficiency. This makes it a highly practical solution for both small-scale and large-scale log parsing.

\subsection{LogPM Benchmark}
\label{sec:logpm}

\subsubsection{Experiment depiction}
LogPM benchmark \cite{logpm} was introduced in response to the limitations of the LogHub benchmark. It distinguishes itself by emphasizing the detection of parameters at the character level. More in-depth, drawing inspiration from the field of computer vision, specifically semantic segmentation, LogPM introduces a new output for log parsers called the "Parameter Mask." This novel approach is accompanied by a new metric, the Parameter Mask Agreement (PMA), designed to compare the produced parsers' parameter masks against the ground truth. Furthermore, LogPM endeavors to better replicate real-world prerequisites for performance evaluation by offering significantly larger datasets than those found in LogHub2k.

In favor of transparency, we merged our implementation of Tipping in the LogPM benchmark into the original publicly available LogPM repository\footnote{\url{https://github.com/M3SOulu/LogPMBenchmark}}.

\subsubsection{Experiment results}

In Table \ref{tab:logpmparacc}, Tipping demonstrates promising performance in the parsing accuracy metric in the LogPM benchmark. Tipping performs better in Android, Linux, and Proxifier datasets. In Zookeeper and HDFS, Tipping achieved the maximum accuracy alongside Lenma and Drain. Accordingly, Tipping performs either better or on par with the best performer in the competition. Furthermore, Tipping achieves the best average accuracy of 0.71, consolidating its place as the best overall performer in the PA metric.

\begin{table}[!ht]
\centering
\caption{An in-depth comparison of Tipping's performance with other leading state-of-the-art parsers in the \textbf{LogPM benchmark}, utilizing \textbf{Parsing Accuracy} as the evaluation metric.}
\label{tab:logpmparacc}
\begin{tabular}{lrrrrr}
\toprule
Dataset & Tipping & Brain & Drain & Spell & Lenma \\
\midrule
HPC & 0.47 & \textbf{0.84} & 0.79 & 0.60 & 0.93 \\ 
Zookeeper & \textbf{1.00} & 0.91 & 0.99 & 0.92 & \textbf{1.00} \\ 
Android & \textbf{0.80} & 0.60 & 0.04 & 0.31 & 0.37 \\ 
Apache & 0.49 & 0.19 & 0.19 & 0.19 & \textbf{0.70} \\ 
Hadoop & 0.74 & 0.86 & 0.72 & 0.84 & \textbf{0.94} \\ 
HDFS & \textbf{1.00} & \textbf{1.00} & 0.72 & 0.72 & 0.83 \\ 
Linux & \textbf{0.92} & 0.90 & 0.87 & 0.87 & \textbf{0.92} \\ 
Openstack & 0.72 & 0.24 & 0.12 & 0.06 & \textbf{0.76} \\ 
Proxifier & \textbf{0.90} & 0.13 & 0.00 & 0.07 & 0.00 \\ 
SSH & 0.04 & 0.69 & \textbf{0.91} & 0.23 & 0.04 \\ 
\midrule
Avg. & \textbf{0.71} & 0.63 & 0.53 & 0.48 & 0.65 \\
\bottomrule
\end{tabular}
\end{table}

In Table \ref{tab:logpmpma}, presenting LogPM's parameter mask agreement scores, Tipping emerges as a leading method with an average parameter mask agreement score of 0.87, showcasing its superior performance across various datasets. This performance is indicative of Tipping's effectiveness in accurately identifying and matching parameter masks, a critical aspect of modern log parsing \cite{logpmbench}. Comparatively, Brain and Lenma each hold an average agreement score of 0.80, indicating a commendable performance but still trailing behind Tipping. Both methods show strengths in certain areas, illustrating their potential in specific contexts. Drain and Spell, with average scores of 0.72 and 0.71, respectively, lag further behind, particularly struggling in datasets like Android.

\begin{table}[!ht]
\centering
\caption{An in-depth comparison of Tipping's performance with other leading state-of-the-art parsers in the \textbf{LogPM benchmark}, utilizing \textbf{Parameter Mask Agreement} as the evaluation metric.}
\label{tab:logpmpma}
\begin{tabular}{lrrrrr}
\toprule
Dataset & Tipping & Brain & Drain & Spell & Lenma \\
\midrule
HPC & \textbf{0.88} & 0.80 & 0.78 & 0.77 & 0.82 \\ 
Zookeeper & \textbf{0.97} & 0.96 & \textbf{0.97} & 0.95 & \textbf{0.97} \\ 
Android & \textbf{0.65} & 0.42 & 0.18 & 0.35 & 0.30 \\ 
Apache & \textbf{0.85} & 0.50 & 0.50 & 0.36 & 0.77 \\ 
Hadoop & 0.81 & 0.89 & 0.82 & 0.89 & \textbf{0.90} \\ 
HDFS & \textbf{1.00} & 0.89 & 0.78 & 0.84 & 0.85 \\ 
Linux & \textbf{0.96} & 0.86 & 0.86 & 0.86 & 0.86 \\ 
Openstack & \textbf{0.92} & 0.91 & 0.74 & 0.70 & \textbf{0.92} \\ 
Proxifier & \textbf{0.95} & 0.79 & 0.62 & 0.62 & 0.48 \\ 
SSH & 0.66 & 0.95 & \textbf{0.98} & 0.79 & 0.89 \\ 
\midrule
Avg. & \textbf{0.87} & 0.80 & 0.72 & 0.71 & 0.80 \\
\bottomrule
\end{tabular}
\end{table}

In Table \ref{tab:logpmspeed}, Tipping performs the fastest in every dataset compared to its competition. The performance gap is more contrasted in larger datasets like HDFS, where Tipping performs 250 times better than the slowest parser (Drain). In terms of the averages, Tipping performs 5 times faster the the second best (Brain).

\begin{table}[!ht]
\centering
\caption{An in-depth comparison of Tipping's performance with other state-of-the-art parsers in the \textbf{LogPM benchmark's speed test}. Other parsers' times are divided by the Tipping time to demonstrate their speed compared to Tipping.}
\label{tab:logpmspeed}
\begin{tabular}{lrrrrr}
\toprule
Dataset & Brain & Drain & Lenma & Spell & Tipping \\
\midrule
Android & 2.58 & 25.89 & 5447.93 & 55.85 & \textbf{1.00} \\
Apache & 5.11 & 1.21 & 68.40 & 16.96 & \textbf{1.00} \\
Hadoop & 4.98 & 25.73 & 27.89 & 145.74 & \textbf{1.00} \\
HDFS & 8.55 & 257.82 & 34.18 & 147.77 & \textbf{1.00} \\
HPC & 3.13 & 1.58 & 15.14 & 41.88 & \textbf{1.00} \\
Linux & 4.75 & 4.36 & 45.58 & 116.52 & \textbf{1.00} \\
Openstack & 6.65 & 2.28 & 48.21 & 124.47 & \textbf{1.00} \\
Proxifier & 6.48 & 2.40 & 33.03 & 19.72 & \textbf{1.00} \\
SSH & 6.75 & 1.47 & 29.53 & 25.45 & \textbf{1.00} \\
Zookeeper & 5.62 & 2.13 & 27.81 & 25.77 & \textbf{1.00} \\
\midrule
Avg. & 5.46 & 32.49 & 577.77 & 72.01 & \textbf{1.00} \\
\bottomrule
\end{tabular}
\end{table}

Overall, Tipping's performance underscores its reliability and capacity to adapt to varied data types, making it a preferred choice for parsing tasks that are similar to LogPM.

\subsection{Hyper-parameter Sensitivity}
\label{sec:hypersens}

\subsubsection{Experiment description}
The previous experiments demonstrate Tipping's performance across various benchmarks. However, its effectiveness could be questioned by the reliance on hyperparameters. In this experiment, we focus on investigating the impact of the threshold parameter, $\theta$, and the inclusion of special patterns (while and black).

To find the sensitivity of each hyperparameter, we repeat the LogHub2k and LogHub2.0 from Sec \ref{sec:logpai} and LogPM from Sec \ref{sec:logpm} with a variety of settings and analyze the results. More specifically, we repeat all three experiments with $\theta$ values in the range of $[0.1, 1.0]$, both in the presence and absence of special tokens.

\begin{figure}[t]
\label{fig:sensloghub2k}
\centering
\includegraphics[width=\linewidth]{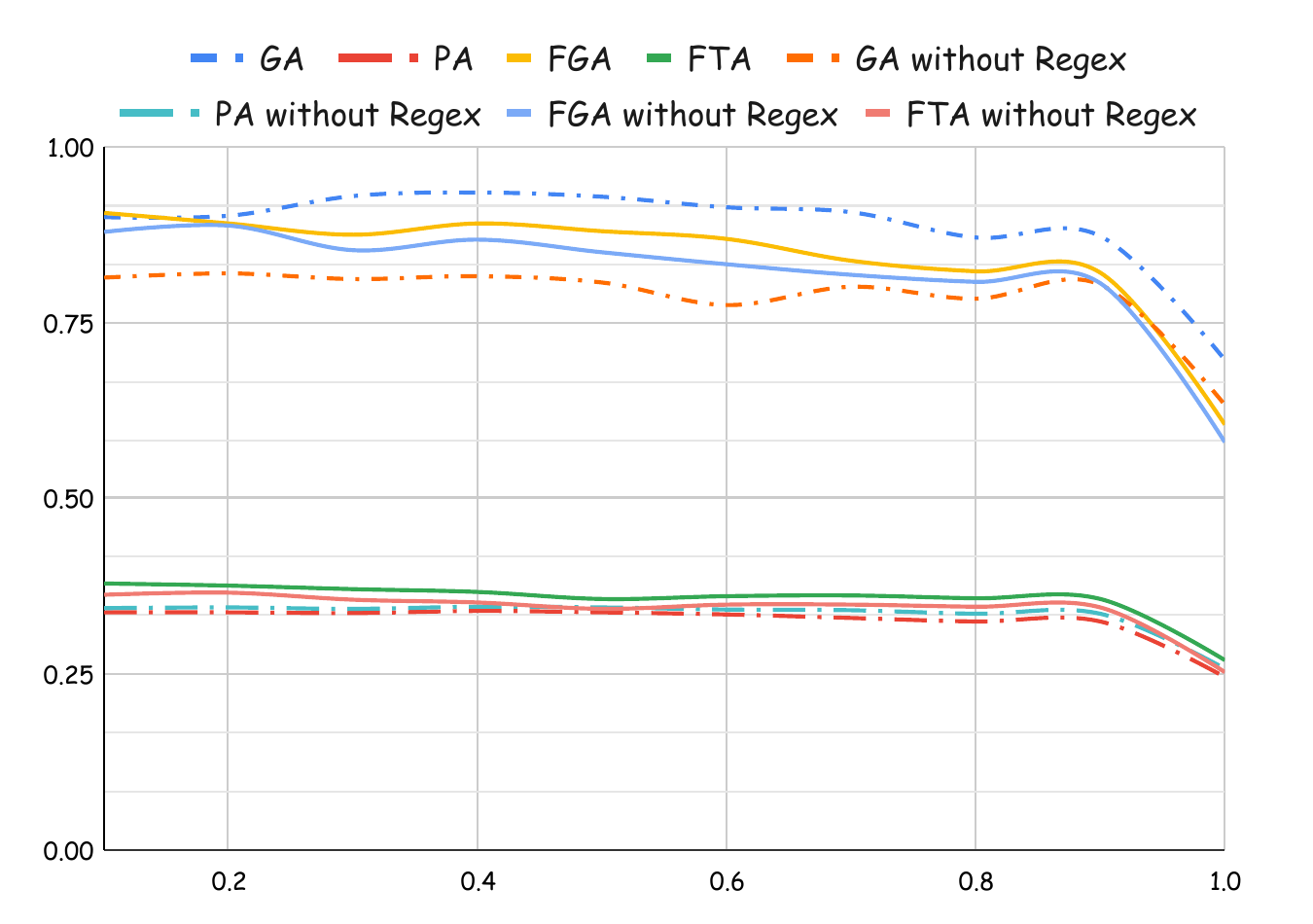}
\centering
\caption{LogHub2k sensitivity experiment for $\theta$ values in the range of $[0.1, 1.0]$ and both with and without black and white special token regexes. The X-axis represents the $\theta$ value, while the Y-axis shows the respective metrics.}
\end{figure}

Figure \ref{fig:sensloghub2k} illustrates the sensitivity test results for LogHub2k. The results indicate that Tipping performs worse in the absence of regexes. Notably, the metrics exhibit consistent behavior in response to $\theta$ variations, regardless of whether regexes are used, with the exception of the extreme high-end. However, the performance gap attributed to regexes is more pronounced for GA than for TA.

\subsubsection{Experiment results}

\begin{figure}[t]
\label{fig:sensloghub2.0}
\centering
\includegraphics[width=\linewidth]{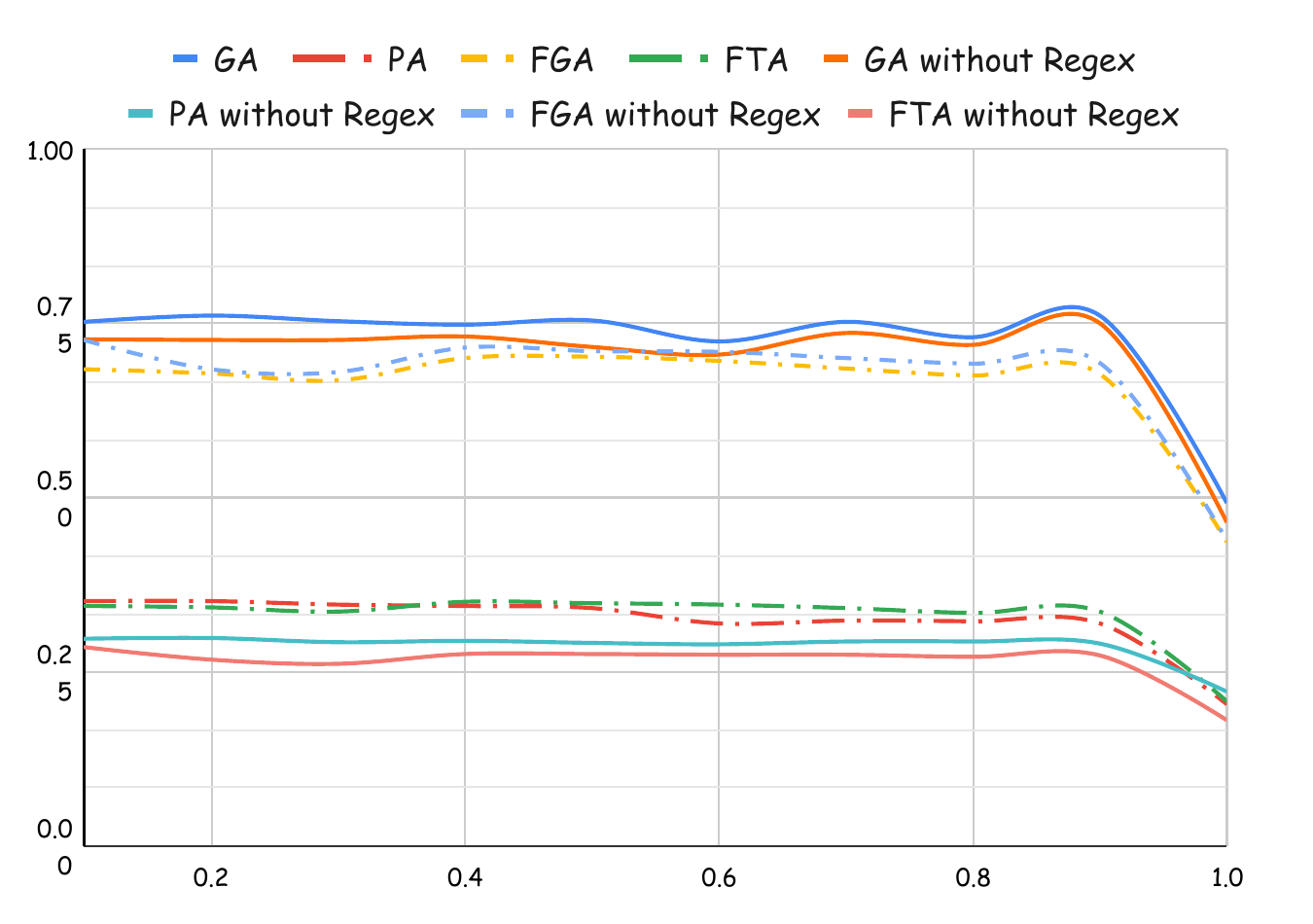}
\centering
\caption{LogHub2.0 sensitivity experiment for $\theta$ values in the range of $[0.1, 1.0]$ and both with and without black and white special token regexes. The X-axis represents the $\theta$ value, while the Y-axis shows the respective metrics.}
\end{figure}

Similarly, Figure \ref{fig:sensloghub2.0} showcases comparable findings for LogHub2.0. While the regex performance gap is less significant, the overall behavioral patterns remain consistent.

\begin{figure}[t]
    \label{fig:senslogpm}
    \centering
    \includegraphics[width=\linewidth]{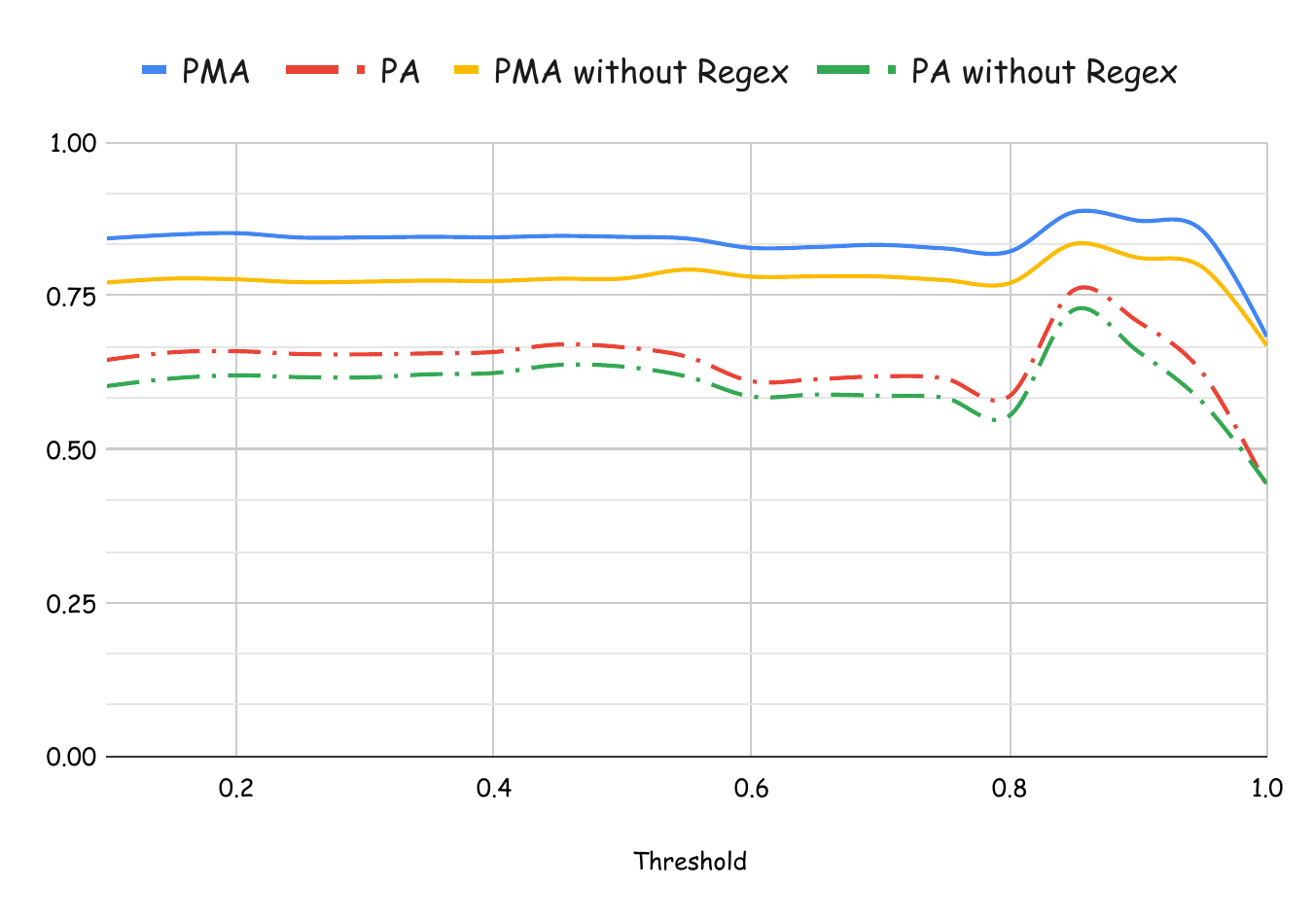}
    \centering
    \caption{LogPM sensitivity experiment for $\theta$ values in the range of $[0.1, 1.0]$ and both with and without black and white special token regexes. The X-axis represents the $\theta$ value, while the Y-axis shows the respective metrics.}
\end{figure}

Lastly, Figure \ref{fig:senslogpm} reveals similar observations for LogPM. Despite differences in metrics, the special tokens' performance gap and the similarity in behavior across different values of $\theta$ follow the same pattern as the previous analysis.

Overall, the observations bear witness to the fact that, except for the extreme ends, despite the possible gains from cherry-picking $\theta$ value and special tokens, the performance of Tipping is not entirely dependent on cherry-picked hyperparameters.

\subsection{LogLead Computational Efficiency Benchmark}
\label{sec:loglead1}
\subsubsection{Experiment depiction}

LogLead\footnote{\url{https://github.com/EvoTestOps/LogLead}} is a tool designed to benchmark log representations and anomaly detection algorithms, such as log parsing results \cite{mantyla2023loglead}. It features integrations with many public datasets and log parsers. LogLead is built with efficiency in mind, achieved by using the Polars\footnote{\url{https://github.com/pola-rs/polars}} library, a notably faster alternative to Pandas\footnote{\url{https://github.com/pandas-dev/pandas}}. As LogLead integrates existing datasets and parsers, it offers an environment to evaluate the computational efficiency of different log parsers over different datasets. Unlike the experiments in the previous sections, no dataset-specific or parser-specific optimizations were made, as they are not available on the LogLead platform.

The scripts we used for our LogLead experiments are  available\footnote{\url{https://github.com/EvoTestOps/LogLead/tree/main/demo/parser_benchmark}}.
For this experiment, we first loaded and preprocessed log data from nine different systems from past studies. This preprocessing involved loading data into Polars data frames and normalizing it using a standard regular expression in the LogLead Enhancer class. The datasets included BGL, Liberty, Spirit, and Thunderbird, originating from real-world high-performance computing usage \cite{bgl}. The Hadoop\cite{hadoopdataset} and HDFS\cite{hdfs} datasets contain logs from experiments conducted on distributed file systems. Additionally, Trainticket \cite{zhou2018fault} and Web-shop\footnote{\url{https://github.com/GoogleCloudPlatform/microservices-demo}} datasets represent experiment logs from modern microservice-based systems \cite{yu2023nezha}. Finally, the Android2 dataset, not to be confused with the Android dataset, originates from long-lasting Android reliability testing runs \cite{mantyla2022pinpointing}.

After preparing the logs, we parsed them with seven log parsers and measured the time. To assess performance differences, we tested all datasets with shares ranging from 1/1024 to 1/1, with each step doubling the amount of data parsed. This arrangement allows us to easily infer whether the parsing time of a particular parser grows linearly, as we would expect the parsing time to double at each step. We conducted the measurements on a local machine (LM) with an Intel® Core™ i7-1365U Processor (two performance cores (max 5.2 GHz) and eight efficiency cores max(3.9 GHz)) with 32Gb of physical memory, and a cloud virtual machine VM with Intel Core Processor (Broadwell, IBRS) CPU 2,4Ghz, 28 cores, and 224GB memory. It should be noted that the LM has more powerful processor cores, while the VM has a higher number of cores available.

Parser implementations for AEL, Brain, and IpLom were taken from \cite{loghubtools}. For Spell\footnote{\url{https://github.com/bave/pyspell}} and LenMa\footnote{\url{https://github.com/keiichishima/templateminer}}, we used the implementations available in the footnotes. For Drain, we used the so-called Drain3\footnote{\url{https://github.com/logpai/Drain3/}} implementation. In the Drain configuration file, we turned off Drain's internal log message normalization via regular expressions and disabled profiling to ensure maximal parsing speed. For other parsers, no such settings were available. 

As our purpose was to investigate the benefits of parallelization, we decided to implement a paralyzation version of the IpLom parser. We named it PlIpLom, Pl being short for Polars. We followed the original IpLom paper and the Python-based source code implementation from \cite{loghubtools}, and we replaced every loop we encountered with native Polars operations if possible. Native Polars operations are known for their efficiency.

\subsubsection{Experiment results}

During our experiments, we set a run-time limit of 10 minutes (600 seconds for each parser on each dataset). Accordingly, if a run exceeded 10 minutes, we allowed it to finish, but no further runs were executed. The remaining results are marked as "Out of Time Limit" (OTL) in the tables from \ref{tab:small_speed} to \ref{tab:liberty_speed_VM}. We also mark crashed runs as "Did Not Finish" (DNF). We conducted a re-run for all crashes; however, if the re-run also failed to complete, we marked it as DNF.

For small and medium datasets, we show only results from parsing full (1/1) datasets. For large data sets, we show fractions of the data as well. The tables showing the fraction results are in our replication package\footnote{https://github.com/M3SOulu/tipping-detailed-result-tables/tree/main/4.4}

We have two small log datasets. The results for the Hadoop dataset (177,592 log lines) on the local machine (LM) and virtual machine (VM) are shown in Table \ref{tab:small_speed}. Tipping, Drain, and PlIplom are the three fastest parsers in both the VM and LM settings. Similar to Hadoop, TrainTicket is a small dataset (272,270 log lines), and its results are documented in Table \ref{tab:small_speed}. In the TrainTicket dataset, Drain is the fastest on the LM, followed by Tipping and Iplom. However, the performance ranking on the VM changes, with Tipping being the fastest, followed by Drain and Iplom.

\begin{table}[]
\caption{Parsing times (in seconds) for parsers across full of small data sets}
\label{tab:small_speed}
\centering
\begin{tabular}{lrrrr}
\toprule
Data & Hadoop & Hadoop & Nezha-TT & Nezha-TT \\
Machine & LM & VM & LM &VM \\
\midrule
AEL & 6.41 & 13.29 & 16.86 & 37.60 \\
Brain & 9.81 & 21.18 & 32.58 & 63.20 \\
Drain & 0.93 & 2.22 & 1.74 & 3.70 \\
Iplom & 2.29 & 5.01 & 6.28 & 13.55 \\
Lenma & 29.54 & 86.78 & 122.89 & 349.02 \\
Pliplom & 1.66 & 2.80 & 9.69 & 14.72 \\
Spell & 47.52 & 105.34 & 78.96 & 164.07 \\
Tipping & 0.22 & 0.36 & 2.01 & 1.11 \\
\bottomrule
\end{tabular}
\end{table}

Next, we have four medium-sized datasets. 
Web-shop, with 3,958,203 log lines, presents its results for the local machine (LM) and virtual machine (VM) in Table \ref{tab:speed_med}. In the LM setting, the top three parsers are PlIplom, Tipping, and Drain, whereas in the VM, the order is Tipping, PlIplom, and Drain. Here, we begin to observe meaningful differences in parsing speed, with the difference between the first and third parsers approximately 9x in LM and 8x in VM.

The results for the BGL dataset, containing 4,747,963 log lines, are displayed in Table \ref{tab:speed_med}. In both computing environments, the top three parsers, in order, are Tipping, PlIplom, and Drain.

The Android2 dataset, with 5,203,599 log lines, shows results in Table \ref{tab:speed_med}. Tipping is the fastest parser in both the VM and LM, while the second and third places alternate between Drain and PlIplom in the VM and LM, respectively.

Finally, the HDFS dataset, which consists of 11,175,629 log lines, is presented in Table \ref{tab:speed_med}. In the LM, PlIplom is the fastest, followed by Tipping and Drain, while in the VM, Tipping leads, followed by PlIplom and Drain.

\begin{table*}[]
\caption{Parsing times (in seconds) for parsers across medium size datasets}
\label{tab:speed_med}
\centering
\begin{tabular}{lrrrrrrrr}
\toprule
Data & Nezha-Shop & Nezha-Shop & BGL & BGL & Android2 & Android2 & HDFS & HDFS \\
Machine & LM & VM & LM & VM & LM & VM & LM & VM \\
\midrule
AEL & 119.08 & 291.96 & 348.72 & 639.22 & 1657.02 & OTL & 338.08 & 727.55 \\
Brain & 191.24 & 357.70 & 169.00 & 354.50 & 210.54 & 468.88 & DNF & 1016.13 \\
Drain & 25.61 & 40.52 & 21.14 & 44.20 & 30.56 & 60.84 & 76.20 & 128.90 \\
Iplom & 39.52 & 106.90 & 43.58 & 100.47 & 57.86 & 119.40 & 119.42 & 256.24 \\
Lenma & OTL & OTL & OTL & OTL & OTL & OTL & OTL & OTL \\
Pliplom & 4.73 & 11.52 & 12.18 & 15.03 & 33.04 & 37.50 & 8.89 & 31.21 \\
Spell & 1556.72 & OTL & OTL & OTL & OTL & OTL & 821.56 & OTL \\
Tipping & 7.69 & 5.89 & 8.86 & 7.35 & 9.44 & 8.82 & 16.44 & 16.60 \\
\bottomrule
\end{tabular}
\end{table*}

Finally, we have three large datasets executed only in the virtual machine (VM), as the local machine's memory size was insufficient. In the VM, none of the parsing executions resulted in fully parsed datasets, as all of the parsers either exceeded the time limit or ran out of memory.

The Thunderbird dataset, with 211,212,192 log lines, shows parsing results in Table \ref{tab:tb_speed_VM}. The fastest parsers, in order, were Tipping, Iplom, and Drain. This order is consistent across the remaining two large datasets. The Liberty and Spirit datasets' results are shown in Tables \ref{tab:liberty_speed_VM} and \ref{tab:sprit_speed_VM}, respectively. Compared to its competitors, the only drawback noted for Tipping is its inability to parse half of the Liberty data, a task that PlIplom manages to accomplish.

\begin{table*}[]
\caption{Parsing times (in seconds) on the \textbf{virtual machine} using \textbf{Thunderbird} dataset for various parsers across different dataset sizes, ranging from 1/1024 to 1/2 of the dataset.}
\label{tab:tb_speed_VM}
\centering
\begin{tabular}{lrrrrrrrrrr}
\toprule
Method & 1/1024 & 1/512 & 1/256 & 1/128 & 1/64 & 1/32 & 1/16 & 1/8 & 1/4 & 1/2 \\
\midrule
AEL & 13.49 & 25.74 & 56.54 & 113.87 & 226.90 & 491.41 & 1013.88 & OTL & OTL & OTL \\
Brain & 17.31 & 37.63 & 88.18 & 179.87 & 350.80 & 640.34 & OTL & OTL & OTL & OTL \\
Drain & 2.48 & 5.18 & 11.45 & 24.03 & 45.45 & 100.54 & 195.61 & 386.25 & 800.11 & OTL \\
%Fiplom & 72.49 & 89.79 & 112.85 & 151.40 & 216.91 & 310.45 & 230.32 & 155.57 & OTL & OTL \\
Iplom & 4.87 & 7.63 & 16.31 & 48.33 & 99.79 & 200.76 & 396.21 & 963.39 & OTL & OTL \\
Lenma & 287.93 & 638.99 & OTL & OTL & OTL & OTL & OTL & OTL & OTL & OTL \\
Pliplom & 3.73 & 5.80 & 7.33 & 12.44 & 20.81 & 34.72 & 74.82 & 142.60 & 322.63 & 720.07 \\
Spell & 446.08 & 1028.88 & OTL & OTL & OTL & OTL & OTL & OTL & OTL & OTL \\
Tipping & 0.97 & 0.91 & 1.62 & 3.44 & 6.46 & 12.64 & 26.88 & 83.50 & 174.77 & 292.95 \\
\bottomrule
\end{tabular}

\end{table*}

\begin{table*}[]
\caption{Parsing times (in seconds) on the \textbf{virtual machine} using \textbf{Spirit} dataset for various parsers across different dataset sizes, ranging from 1/1024 to 1/4 of the dataset.}
\label{tab:sprit_speed_VM}
\centering
\begin{tabular}{lrrrrrrrrrr}
\toprule
Method & 1/1024 & 1/512 & 1/256 & 1/128 & 1/64 & 1/32 & 1/16 & 1/8 & 1/4 \\
\midrule
AEL & 18.14 & 35.69 & 72.63 & 146.75 & 326.59 & 671.77 & OTL & OTL & OTL \\
Brain & 27.04 & 56.41 & 126.17 & 253.24 & 556.99 & 1119.26 & OTL & OTL & OTL \\
Drain & 3.37 & 7.32 & 13.93 & 27.02 & 61.53 & 122.03 & 235.50 & 411.58 & DNF \\
Fiplom & 69.21 & 93.98 & 149.81 & 265.75 & 484.10 & 869.52 & OTL & OTL & OTL \\
Iplom & 5.79 & 10.68 & 31.64 & 65.18 & 138.55 & 291.27 & 514.12 & 1100.10 & OTL \\
Lenma & 207.35 & 497.37 & 1487.78 & OTL & OTL & OTL & OTL & OTL & OTL \\
Pliplom & 3.45 & 5.57 & 11.14 & 19.48 & 38.72 & 72.27 & 170.26 & 378.54 & 984.92 \\
Spell & 417.94 & 1006.59 & OTL & OTL & OTL & OTL & OTL & OTL & OTL \\
Tipping & 1.38 & 1.11 & 2.09 & 3.79 & 7.56 & 16.85 & 31.08 & 104.33 & 119.07 \\
\bottomrule
\end{tabular}

\end{table*}

\begin{table*}[]
\caption{Parsing times (in seconds) on the \textbf{virtual machine} using \textbf{Liberty} dataset for various parsers across different dataset sizes, ranging from 1/1024 to 1/2 of the dataset.}
\label{tab:liberty_speed_VM}
\centering
\begin{tabular}{lrrrrrrrrrr}
\toprule
Parser & 1/1024 & 1/512 & 1/256 & 1/128 & 1/64 & 1/32 & 1/16 & 1/8 & 1/4 & 1/2 \\
\midrule
AEL & 18.11 & 32.47 & 66.77 & 130.24 & 265.87 & 553.11 & 1247.87 & OTL & OTL & OTL \\
Brain & 17.67 & 35.54 & 77.04 & 168.11 & 350.12 & 675.11 & OTL & OTL & OTL & OTL \\
Drain & 2.37 & 4.99 & 10.10 & 21.48 & 46.38 & 98.69 & 182.59 & 347.35 & 707.99 & OTL \\
Fiplom & 38.95 & 59.50 & 84.10 & 138.30 & 213.22 & 369.22 & 658.94 & OTL & OTL & OTL \\
Iplom & 4.27 & 7.77 & 15.96 & 47.13 & 90.16 & 185.26 & 378.99 & 824.79 & OTL & OTL \\
Lenma & 149.24 & 385.89 & 999.13 & OTL & OTL & OTL & OTL & OTL & OTL & OTL \\
Pliplom & 2.20 & 3.59 & 5.90 & 11.96 & 23.17 & 39.74 & 80.59 & 171.82 & 230.52 & 478.18 \\
Spell & 218.75 & 563.70 & 1503.56 & OTL & OTL & OTL & OTL & OTL & OTL & OTL \\
Tipping & 0.80 & 0.81 & 1.85 & 3.55 & 6.20 & 12.57 & 24.62 & 51.10 & 116.22 & DNF \\
\bottomrule
\end{tabular}
\end{table*}

% \begin{figure}
%     \centering
%     \includegraphics[width=1\linewidth]{output.png}
%     \caption{Enter Caption}
%     \label{fig:enter-label}
% \end{figure}

\subsubsection{Summary}
%To summarize, we did speed test experiment in two computing platforms and with 9 dataset, altogether 15 experiments as we were unable to run 3 largest dataset in our local machine. We find that tipping is the fastest in 13 out of the 15 experiments. In two LM machine experiments, Tipping becomes second to PlIpLom but is faster in a VM with more CPU cores, but the cores themselves are slower. This is an indication of Tipping's superior parallelization capability. The second fastest parser is PlIpLom, scoring two first places, 9 second places, and three third places. Drain is the third fastest with no victories, 5 second places, and 10 third places. The difference between PlIpLom and Drain is that PlIpLom tends to be faster than Drain in medium and large datasets, while Drain is faster in small datasets. 

To summarize, we conducted speed test experiments on two computing platforms with nine datasets, totaling 15 experiments, as we could not run the three most enormous datasets on our local machine. We found that Tipping was the fastest parser in 12 out of the 15 experiments. In two experiments on the local machine (HDFS and WebShop), Tipping was second to PlIplom but then performed faster on the virtual machine, which has more CPU cores, though the cores are slower. A similar result was observed in the local machine with TrainTicket, where Drain was the fastest. This provides evidence of Tipping's superior parallelization capability. The second fastest parser was PlIplom, which secured two first places. Drain was the third fastest, with one first place. The notable difference between PlIplom and Drain is that PlIplom tends to be faster than Drain in medium and large datasets, while Drain performs better in small datasets.

%Drain
%1:1
%2:1111-1-
%3:-1111111-111

%PLiplom
% 1:-11- (LM HDFS, LM WebShop)
% 2:1-11111-111
% 3:11-1-
% O:11--

Investigating Tipping's parallelization capability further, we compared how Tipping performs with itself when run on the same dataset in LM and VM. 
Remember that VM has multiple, slower processors, while LM has fewer but faster processors.  We found that Tipping is faster in the VM in four of the six datasets. In the HDFS dataset, Tipping is faster in the LM by a small margin of 1\%. Furthermore, Tipping performs better in the LM on the Hadoop dataset, which is considered a small dataset with parsing times less than 0.5 seconds, indicating that parallelization does not pay off in such small datasets.

The fact that Tipping is faster in the VM distinguishes it from all other parsers, as all other parsers consistently show slower performance in the VM compared to the LM. For example, Drain, which has been the benchmark parser for years and was the overall third fastest parser in our tests, shows significantly slower performance in the VM across various datasets. Specifically, Drain is 139\%, 126\%, 58\%, 110\%, 99\%, and 69\% slower in the VM than in the LM for the Hadoop, TrainTicket, Web-shop, BGL, Android, and HDFS datasets, respectively. This contrast underscores Tipping's superior parallelization capabilities when deployed in environments with more CPU cores.

\subsection{Anomaly Detection benchmark with LogLead}
\label{sec:loglead2}
\subsubsection{Experiment depiction}
In our final experiment, we examine how the parsing output of different parsers affects the downstream task of anomaly detection in software logs. We used the LogLead tool described in Section \ref{sec:loglead1} to carry out the experiments. 

Past work provides somewhat mixed results on whether the accuracy of log parsing impacts the performance of anomaly detection \cite{parsereffect}. Nevertheless, in a real-world scenario, the accuracy of the end goals is an important property of analysis pipelines, as ideally, one chooses components based on parsing performance and the end goal's accuracy.

We used 3 datasets that we were able to fully parse: Hadoop, HDFS, and BGL. The remaining supercomputer datasets, Thunderbird, Liberty, and Spirit, were excluded due to their size. Trainticket and Web-shop are multimodal datasets consisting of not only logs but also traces and metrics, which must be analyzed to achieve decent performance on those datasets. Finally, our Android2 dataset contained only 2 anomaly instances, making it impossible to use for anomaly detection investigations. 

We tested the top three fastest parsers from the previous experiment: Tipping, PlIpLom, and Drain alongside Brain, as it had shown good accuracy in our first and second experiments.

We tested a combination of supervised and unsupervised algorithms to determine if parsing results impact anomaly detection accuracy. We used the Decision Tree (su-DT), Linear Support-Vector Machine (su-SVM), Logistic Regression (su-LR), Random Forest (su-RF), and Extreme Gradient Boosting Tree (su-XGB) for supervised methods, and K-means clustering (us-KM) and Isolation Forest (us-IF) for unsupervised methods. Additionally, we tested a custom-made unsupervised model called RarityModel (us-RM) \cite{nyyssola2023efficiency} that utilizes the negative logarithm to measure the rarity of individual log events. This model proposes that the rarer a log event, the higher the likelihood of an anomaly.

We used two accuracy measures: $F_1$ score and AUC-ROC. The $F_1$ score indicates how well the model detects anomalies with a fixed (optimal) threshold. The AUC-ROC score shows how well the model detects anomalies if one considers all possible thresholds. The AUC-ROC score is important for practical usage, as higher precision or recall may be required in certain use cases.

Our training and testing procedure varied between datasets as they are very different in nature. HDFS and Hadoop are experimental datasets, meaning that researchers had run the systems in a lab while injecting anomalies in an arbitrary order. Thus, there is no real-world chronological order, and sequences from those datasets can be placed in random order when testing anomaly detection performance. However, random order cannot be used for data originating from real-world usage, as one cannot use future data to predict past performance. This is the case for BGL, which comes from production use over 200 days.

For Hadoop, we randomized data, split the data to 50-50 for training and testing, repeated the procedure 100 times, and computed the average of $F_1$ score and AUC-ROC for each log parser and machine learning algorithm. Hadoop is a fairly small dataset, so a 50-50 split was needed to get decent performance. HDFS, however, suffers from the opposite problem, i.e., results tend to be too good if the entire dataset is used. Therefore, we used only random 1\% of data and repeated this procedure 100 times. For each 1\%, we did ten random splits while using 10\% data for training and 90\% for testing. This resulted in a far lower $F_1$ score than prior works but allowed us to better investigate parser differences. For BGL, we maintained chronological order and tested 10 different train-test splits (5-95, 15-85, 25-75, 35-65, 45-55, 55-45, 65-35, 75-25, 85-15, 95-5) and computed the average. 

% For Hadoop, we randomized the data, split the data 50-50 for training and testing, and repeated the procedure 100 times. In Hadoop data, anomalies are the majority, so we flipped the target variable to predict non-anomaly minority cases. Then, we computed the average $F_1$ score and AUC-ROC scores for each log parser and machine learning algorithm pair. Hadoop is a fairly small dataset, so a 50-50 split was needed to achieve decent performance. HDFS, however, suffers from the opposite problem, i.e., results tend to be too good if the full dataset is used. Therefore, we sampled only a random 1\% of the data and repeated this procedure 100 times. For each 1\% dataset, we did ten random splits while using 10\% of the data for training and 90\% for testing. This resulted in far lower $F_1$ scores compared to prior works but allowed us to better investigate differences between parsers. For BGL, we maintained chronological order and tested 10 different train-test splits (5-95, 15-85, 25-75, 35-65, 45-55, 55-45, 65-35, 75-25, 85-15, 95-5) and computed the average.

\subsubsection{Experiment results}
Hadoop results are presented in Tables \ref{tab:ad-hadoop-aucroc} and \ref{tab:ad-hadoop-$F_1$ score}. PlIpLom achieves the best results in terms of both $F_1$ score and AUC-ROC, followed by Tipping, while the last place alternates between Drain and Brain. The difference between the winner, PlIpLom, and the last parser in the AUC-ROC average is 0.016 (0.612 - 0.596) and in $F_1$ score 0.024 (0.337 - 0.313). PlIpLom's top performance stems from its results with the SVM model, where it significantly outperforms the last parser with an AUC-ROC of 0.825 compared to 0.750 and an $F_1$ score of 0.574 compared to 0.377. The differences between PlIpLom and other parsers are negligible with all other machine learning models.

% \begin{table}[]
% \caption{Anomaly detection data:Hadoop, score:auc-roc, metric:median, data-proportion:1/1, data-rows:177592, redraws:1, train-test-repeats:100, test-fraction: 0.5, chono-order:not chronological order, normalized:normalized}
% \centering
% \begin{tabular}{llllll}
% \hline
% Parser & Brain & Drain & PlIpLom & Tipping & Avg. \\
% Model &  &  &  &  &  \\
% \midrule
% su-DT & 0.702 & 0.725 & 0.710 & 0.732 & 0.717 \\
% su-SVM & 0.737 & 0.742 & 0.848 & 0.778 & 0.776 \\
% su-LR & 0.745 & 0.742 & 0.756 & 0.758 & 0.750 \\
% su-RF & 0.839 & 0.847 & 0.852 & 0.849 & 0.847 \\
% su-XGB & 0.817 & 0.812 & 0.835 & 0.827 & 0.822 \\
% us-IF & 0.282 & 0.282 & 0.275 & 0.284 & 0.281 \\
% us-KM & 0.432 & 0.423 & 0.419 & 0.428 & 0.426 \\
% us-RM & 0.244 & 0.243 & 0.244 & 0.252 & 0.246 \\
% Avg. & 0.600 & 0.602 & 0.617 & 0.613 & 0.608 \\
% \hline
% \end{tabular}
% \end{table}

\begin{table}[]
\caption{AUC-ROC of anomaly detection on Hadoop
%, score:auc-roc, metric:mean, data-proportion:1/1, data-rows:177592, redraws:1, train-test-repeats:100, test-fraction: 0.5, chono-order:not chronological order, normalized:normalized
}
\label{tab:ad-hadoop-aucroc}
\centering
\begin{tabular}{lrrrrr}
\toprule
Model & Brain & Drain & PlIpLom & Tipping & Avg. \\
\midrule
% Model &  &  &  &  &  \\
su-DT & 0.692 & 0.703 & 0.697 & 0.716 & 0.702 \\
su-SVM & 0.740 & 0.740 & 0.825 & 0.768 & 0.768 \\
su-LR & 0.735 & 0.733 & 0.745 & 0.748 & 0.740 \\
su-RF & 0.836 & 0.835 & 0.842 & 0.839 & 0.838 \\
su-XGB & 0.808 & 0.807 & 0.829 & 0.825 & 0.817 \\
us-IF & 0.283 & 0.276 & 0.267 & 0.287 & 0.278 \\
us-KM & 0.442 & 0.432 & 0.442 & 0.443 & 0.440 \\
us-RM & 0.242 & 0.243 & 0.244 & 0.250 & 0.245 \\
\midrule
Avg. & 0.597 & 0.596 & 0.612 & 0.610 & 0.604 \\
\bottomrule
\end{tabular}
\end{table}

% \begin{table}[]
% \caption{Anomaly detection data:Hadoop, score:$F_1$ score, metric:median, data-proportion:1/1, data-rows:177592, redraws:1, train-test-repeats:100, test-fraction: 0.5, chono-order:not chronological order, normalized:normalized}
% \centering
% \begin{tabular}{llllll}
% \hline
% Parser & Brain & Drain & PlIpLom & Tipping & Avg. \\
% Model &  &  &  &  &  \\
% \midrule
% su-DT & 0.500 & 0.513 & 0.500 & 0.530 & 0.511 \\
% su-SVM & 0.400 & 0.400 & 0.571 & 0.444 & 0.454 \\
% su-LR & 0.429 & 0.429 & 0.429 & 0.437 & 0.431 \\
% su-RF & 0.444 & 0.462 & 0.444 & 0.500 & 0.463 \\
% su-XGB & 0.500 & 0.500 & 0.500 & 0.500 & 0.500 \\
% us-IF & 0.000 & 0.000 & 0.000 & 0.121 & 0.030 \\
% us-KM & 0.000 & 0.000 & 0.000 & 0.000 & 0.000 \\
% us-RM & 0.312 & 0.312 & 0.312 & 0.312 & 0.312 \\
% Avg. & 0.323 & 0.327 & 0.345 & 0.356 & 0.338 \\
% \hline
% \end{tabular}
% \end{table}

\begin{table}[]
\caption{$F_1$ score of anomaly detection on Hadoop
%Anomaly detection data:Hadoop, score:$F_1$ score, metric:mean, data-proportion:1/1, data-rows:177592, redraws:1, train-test-repeats:100, test-fraction: 0.5, chono-order:not chronological order, normalized:normalized
}
\label{tab:ad-hadoop-$F_1$ score}
\centering
\begin{tabular}{lrrrrr}
\toprule
Model & Brain & Drain & PlIpLom & Tipping & Avg. \\
% Model &  &  &  &  &  \\
\midrule
su-DT & 0.464 & 0.489 & 0.470 & 0.503 & 0.481 \\
su-SVM & 0.383 & 0.377 & 0.574 & 0.431 & 0.441 \\
su-LR & 0.396 & 0.398 & 0.409 & 0.401 & 0.401 \\
su-RF & 0.414 & 0.440 & 0.432 & 0.445 & 0.433 \\
su-XGB & 0.458 & 0.461 & 0.459 & 0.477 & 0.464 \\
us-IF & 0.056 & 0.080 & 0.072 & 0.106 & 0.078 \\
us-KM & 0.000 & 0.000 & 0.000 & 0.000 & 0.000 \\
us-RM & 0.335 & 0.335 & 0.335 & 0.335 & 0.335 \\
\midrule
Avg. & 0.313 & 0.322 & 0.344 & 0.337 & 0.329 \\
\bottomrule
\end{tabular}
\end{table}

Tables \ref{tab:ad-hdfs-aucroc} and \ref{tab:ad-hdfs-$F_1$ score} demonstrate the HDFS results, showing no meaningful differences between parsers' average $F_1$ scores or AUC-ROCs. However, when examining individual machine learning models and parsers pairs, differences in terms of $F_1$ score are observed (see Table \ref{tab:ad-hdfs-$F_1$ score}). Brain shows slightly weaker performance with LR and RF compared to the rest but compensates with better performance using the IF model. The notable differences between machine learning algorithms suggest that the setup should allow for detecting differences between parsers if there are any. 

% \begin{table}[]
% \caption{Anomaly detection data:Hdfs, score:auc-roc, metric:median, data-proportion:1/100, data-rows:111942, redraws:100, train-test-repeats:10, test-fraction: 0.9, chono-order:not chronological order, normalized:not normalized}
% \centering
% \begin{tabular}{llllll}
% \hline
% Parser & Brain & Drain & PlIpLom & Tipping & Avg. \\
% Model &  &  &  &  &  \\
% \midrule
% su-DT & 0.957 & 0.959 & 0.960 & 0.959 & 0.959 \\
% su-SVM & 0.963 & 0.964 & 0.963 & 0.964 & 0.964 \\
% su-LR & 0.956 & 0.956 & 0.956 & 0.957 & 0.956 \\
% su-RF & 0.982 & 0.982 & 0.982 & 0.982 & 0.982 \\
% su-XGB & 0.945 & 0.945 & 0.945 & 0.945 & 0.945 \\
% us-IF & 0.956 & 0.955 & 0.956 & 0.955 & 0.955 \\
% us-KM & 0.843 & 0.844 & 0.844 & 0.847 & 0.845 \\
% us-RM & 0.598 & 0.598 & 0.598 & 0.598 & 0.598 \\
% Avg. & 0.900 & 0.901 & 0.901 & 0.901 & 0.901 \\
% \hline
% \end{tabular}
% \end{table}

\begin{table}[]
\caption{AUC-ROC of anomaly detection on HDFS
%Anomaly detection data:Hdfs, score:auc-roc, metric:mean, data-proportion:1/100, data-rows:111942, redraws:100, train-test-repeats:10, test-fraction: 0.9, chono-order:not chronological order, normalized:not normalized
}
\label{tab:ad-hdfs-aucroc}
\centering
\begin{tabular}{lrrrrr}
\toprule
Model & Brain & Drain & PlIpLom & Tipping & Avg. \\
% Model &  &  &  &  &  \\
\midrule
su-DT & 0.950 & 0.951 & 0.951 & 0.951 & 0.951 \\
su-SVM & 0.944 & 0.947 & 0.946 & 0.947 & 0.946 \\
su-LR & 0.947 & 0.946 & 0.946 & 0.947 & 0.947 \\
su-RF & 0.975 & 0.975 & 0.975 & 0.975 & 0.975 \\
su-XGB & 0.915 & 0.915 & 0.915 & 0.915 & 0.915 \\
us-IF & 0.951 & 0.951 & 0.951 & 0.951 & 0.951 \\
us-KM & 0.842 & 0.843 & 0.843 & 0.844 & 0.843 \\
us-RM & 0.608 & 0.608 & 0.608 & 0.608 & 0.608 \\
\midrule
Avg. & 0.891 & 0.892 & 0.892 & 0.892 & 0.892 \\
\bottomrule
\end{tabular}
\end{table}

% \begin{table}[]
% \caption{Anomaly detection data:Hdfs, score:$F_1$ score, metric:median, data-proportion:1/100, data-rows:111942, redraws:100, train-test-repeats:10, test-fraction: 0.9, chono-order:not chronological order, normalized:not normalized}
% \centering
% \begin{tabular}{llllll}
% \hline
% Parser & Brain & Drain & PlIpLom & Tipping & Avg. \\
% Model &  &  &  &  &  \\
% \midrule
% su-DT & 0.936 & 0.937 & 0.937 & 0.937 & 0.937 \\
% su-SVM & 0.907 & 0.910 & 0.910 & 0.910 & 0.909 \\
% su-LR & 0.645 & 0.647 & 0.646 & 0.647 & 0.646 \\
% su-RF & 0.909 & 0.911 & 0.910 & 0.910 & 0.910 \\
% su-XGB & 0.875 & 0.874 & 0.875 & 0.875 & 0.875 \\
% us-IF & 0.420 & 0.418 & 0.413 & 0.414 & 0.416 \\
% us-KM & 0.000 & 0.000 & 0.000 & 0.000 & 0.000 \\
% us-RM & 0.105 & 0.107 & 0.107 & 0.107 & 0.107 \\
% Avg. & 0.599 & 0.600 & 0.600 & 0.600 & 0.600 \\
% \hline
% \end{tabular}
% \end{table}

\begin{table}[]
\caption{$F_1$ score of anomaly detection on HDFS
%Anomaly detection data:Hdfs, score:$F_1$ score, metric:mean, data-proportion:1/100, data-rows:111942, redraws:100, train-test-repeats:10, test-fraction: 0.9, chono-order:not chronological order, normalized:not normalized
}
\label{tab:ad-hdfs-$F_1$ score}
\centering
\begin{tabular}{lrrrrr}
\toprule
Model & Brain & Drain & PlIpLom & Tipping & Avg. \\
% Model &  &  &  &  &  \\
\midrule
su-DT & 0.928 & 0.929 & 0.929 & 0.929 & 0.929 \\
su-SVM & 0.900 & 0.902 & 0.902 & 0.902 & 0.901 \\
su-LR & 0.644 & 0.647 & 0.647 & 0.648 & 0.646 \\
su-RF & 0.897 & 0.899 & 0.900 & 0.900 & 0.899 \\
su-XGB & 0.836 & 0.834 & 0.836 & 0.836 & 0.836 \\
us-IF & 0.438 & 0.434 & 0.429 & 0.429 & 0.433 \\
us-KM & 0.000 & 0.000 & 0.000 & 0.000 & 0.000 \\
us-RM & 0.122 & 0.125 & 0.125 & 0.125 & 0.124 \\
\midrule
Avg. & 0.596 & 0.596 & 0.596 & 0.596 & 0.596 \\
\bottomrule
\end{tabular}
\end{table}

BGL results are in Tables \ref{tab:ad-bgl-aucroc} and \ref{tab:ad-bgl-$F_1$ score}. Tipping is the winner while the margin to the last parser in AUC-ROC is 0.068 (0.696 - 0.628) and in $F_1$ score 0.104 (0.329 - 0.225). Tipping is superior with all supervised algorithms and IF from the unsupervised algorithms. With unsupervised algorithms, Brain is better than Tipping with RM and is on par with KM.

% \begin{table}[]
% \caption{Anomaly detection data:Bgl5-95, score:auc-roc, metric:median, data-proportion:1/1, data-rows:4713493, anos:348460 redraws:1, train-test-repeats:1, test-fraction: [0.05, 0.15, 0.25, 0.35, 0.45, 0.55, 0.65, 0.75, 0.85, 0.95], chono-order:True, normalize:True}
% \centering
% \begin{tabular}{llllll}
% \hline
% Parser & Brain & Drain & PlIpLom & Tipping & Avg. \\
% Model &  &  &  &  &  \\
% \midrule
% su-DT & 0.629 & 0.629 & 0.629 & 0.771 & 0.664 \\
% su-SVM & 0.826 & 0.824 & 0.825 & 0.903 & 0.844 \\
% su-LR & 0.826 & 0.824 & 0.825 & 0.903 & 0.844 \\
% su-RF & 0.629 & 0.629 & 0.629 & 0.839 & 0.681 \\
% su-XGB & 0.769 & 0.768 & 0.777 & 0.838 & 0.788 \\
% us-IF & 0.276 & 0.295 & 0.294 & 0.369 & 0.309 \\
% us-KM & 0.545 & 0.496 & 0.522 & 0.558 & 0.530 \\
% us-RM & 0.421 & 0.423 & 0.423 & 0.428 & 0.424 \\
% Avg. & 0.615 & 0.611 & 0.616 & 0.701 & 0.636 \\
% \hline
% \end{tabular}
% \end{table}

\begin{table}[]
\caption{AUC-ROC of anomaly detection on BGL
%Anomaly detection data:Bgl5-95, score:auc-roc, metric:mean, data-proportion:1/1, data-rows:4713493, anos:348460 redraws:1, train-test-repeats:1, test-fraction: [0.05, 0.15, 0.25, 0.35, 0.45, 0.55, 0.65, 0.75, 0.85, 0.95], chono-order:True, normalize:True
}
\label{tab:ad-bgl-aucroc}
\centering
\begin{tabular}{lrrrrr}
\toprule
Model & Brain & Drain & PlIpLom & Tipping & Avg. \\
% Model &  &  &  &  &  \\
\midrule
su-DT & 0.638 & 0.639 & 0.639 & 0.749 & 0.667 \\
su-SVM & 0.842 & 0.837 & 0.844 & 0.907 & 0.858 \\
su-LR & 0.842 & 0.841 & 0.845 & 0.908 & 0.859 \\
su-RF & 0.638 & 0.639 & 0.639 & 0.782 & 0.675 \\
su-XGB & 0.793 & 0.789 & 0.797 & 0.854 & 0.808 \\
us-IF & 0.334 & 0.396 & 0.364 & 0.400 & 0.373 \\
us-KM & 0.541 & 0.457 & 0.491 & 0.541 & 0.507 \\
us-RM & 0.451 & 0.425 & 0.425 & 0.426 & 0.432 \\
\midrule
Avg. & 0.635 & 0.628 & 0.631 & 0.696 & 0.647 \\
\bottomrule
\end{tabular}
\end{table}

% \begin{table}[]
% \caption{Anomaly detection data:Bgl5-95, score:$F_1$ score, metric:median, data-proportion:1/1, data-rows:4713493, anos:348460 redraws:1, train-test-repeats:1, test-fraction: [0.05, 0.15, 0.25, 0.35, 0.45, 0.55, 0.65, 0.75, 0.85, 0.95], chono-order:True, normalize:True}
% \centering
% \begin{tabular}{llllll}
% \hline
% Parser & Brain & Drain & PlIpLom & Tipping & Avg. \\
% Model &  &  &  &  &  \\
% \midrule
% su-DT & 0.405 & 0.406 & 0.402 & 0.698 & 0.478 \\
% su-SVM & 0.305 & 0.406 & 0.402 & 0.698 & 0.453 \\
% su-LR & 0.271 & 0.273 & 0.274 & 0.356 & 0.294 \\
% su-RF & 0.405 & 0.406 & 0.402 & 0.698 & 0.478 \\
% su-XGB & 0.270 & 0.272 & 0.274 & 0.356 & 0.293 \\
% us-IF & 0.000 & 0.000 & 0.000 & 0.000 & 0.000 \\
% us-KM & 0.000 & 0.000 & 0.000 & 0.000 & 0.000 \\
% us-RM & 0.001 & 0.007 & 0.007 & 0.007 & 0.006 \\
% Avg. & 0.207 & 0.221 & 0.220 & 0.352 & 0.250 \\
% \hline
% \end{tabular}
% \end{table}

\begin{table}[]
\caption{$F_1$ score of anomaly detection on BGL
%Anomaly detection data:Bgl5-95, score:$F_1$ score, metric:mean, data-proportion:1/1, data-rows:4713493, anos:348460 redraws:1, train-test-repeats:1, test-fraction: [0.05, 0.15, 0.25, 0.35, 0.45, 0.55, 0.65, 0.75, 0.85, 0.95], chono-order:True, normalize:True
}
\centering
\label{tab:ad-bgl-$F_1$ score}
\begin{tabular}{lrrrrr}
\toprule
Model & Brain & Drain & PlIpLom & Tipping & Avg. \\
% Model &  &  &  &  &  \\
\midrule
su-DT & 0.373 & 0.377 & 0.375 & 0.559 & 0.421 \\
su-SVM & 0.336 & 0.377 & 0.375 & 0.559 & 0.412 \\
su-LR & 0.339 & 0.343 & 0.339 & 0.483 & 0.376 \\
su-RF & 0.373 & 0.377 & 0.375 & 0.559 & 0.421 \\
su-XGB & 0.332 & 0.337 & 0.333 & 0.463 & 0.366 \\
us-IF & 0.002 & 0.002 & 0.000 & 0.000 & 0.001 \\
us-KM & 0.000 & 0.000 & 0.000 & 0.000 & 0.000 \\
us-RM & 0.047 & 0.010 & 0.011 & 0.012 & 0.020 \\
\midrule
Avg. & 0.225 & 0.228 & 0.226 & 0.329 & 0.252 \\
\bottomrule
\end{tabular}
\end{table}

\subsubsection{Summary}
We received mixed results when testing four parsers in three datasets with 8 different algorithms. For the HDFS dataset, there was no difference between parsers, resulting in an astonishingly equal four-way tie. In Hadoop, PlIplon was the best, but its victory mainly stemmed from being widely superior with one algorithm (SVM), while with other machine learning models, the differences were negligible. In BGL, Tipping was superior with all supervised models and one unsupervised model, but with two unsupervised models, Tipping was no better than the others.

Past work \cite{parsereffect} has proposed that the number of templates a parser generates is negatively correlated with the anomaly detection capability of the parser. We checked the number of events the four parsers produced, as shown in Table
\ref{tab:n-events-in-ad}, and found mixed results. In HDFS, Drain produces 43\% more events than Tipping, but there is no difference in anomaly detection capability in that dataset. In Hadoop, PlIplom was the best parser, but it did not produce the smallest amount of events. This theory only holds in the BGL dataset, where Tipping produced the smallest number of events and was the best parser.

Table \ref{tab:n-events-in-ad} also compares the "correct" number of templates from two sources produced by humans on the full dataset \cite{logpm, landauer2023critical}. Event numbers from the sources do not match each other, indicating that it is hard to arrive at the same result, even when we agree on the concept of templates and parameters on a higher level. Using those numbers suggests that all parsers are in the right ballpark regarding HDFS and Hadoop. However, regarding BGL, many parsers seem to produce too many templates. Thus, this would bring forward a new hypothesis that the number of templates decreases the anomaly detection accuracy only if the number deviates too much from the "correct" number of templates.

\begin{table}[]
\caption{Number of events produced}
\centering
\label{tab:n-events-in-ad}
\begin{tabular}{lllllll}
\toprule
Dataset & Brain & Drain & PlIpLom & Tip & \cite{landauer2023critical} & \cite{logpm} \\
% Data &  &  &  &  &  \\
\midrule
Hadoop & 306 & 292 & 267 & 233 & 312* & 253\\
HDFS & 44 & 53 & 46 & 37 & 33  & 53\\
BGL & 643 & 850 & 1018 & 302 & 394  & NA\\
\bottomrule
\multicolumn{7}{l}{* We subtracted event \#37 that described stack traces}
\end{tabular}
\end{table}

\section{Threats to Validity}
\subsection{Internal Validity}
Our experiments were conducted on specific hardware configurations, including machines with multiple CPU cores. The performance gap between Tipping and other parsers might not be as wide as our experiment under different hardware settings, especially when the target machine lacks in quantity of CPU cores. However, this situation seems not to be prevalent in today's age of technology.

Although Tipping's and the benchmark's implementations undergo rigorous testing, the outcomes may inadvertently indicate a bug which may invalidate evaluations.

We allocated no more than five minutes to parameter tuning per dataset in experiments in Sections \ref{sec:logpai} and \ref{sec:logpm}. However, it might not reflect a realistic timeframe, as the authors are knowledgeable and familiar with both the datasets and the algorithm.

\subsection{External Validity}
The datasets utilized in our study were chosen to represent a broad range of real-world software systems. However, the open public datasets used are not sourced from the software's latest version and may not encompass all possibilities. Furthermore, our experiments were conducted in controlled environments, which may not perfectly replicate the complexities and variabilities of real-world log data. Therefore, while our results are indicative of Tipping’s performance across diverse conditions, they may not fully generalize to all log parsing contexts.

\section{Conclusion}
In this paper, we introduced a novel log parsing algorithm named "Tipping," which utilizes a combination of rule-based tokenizers, interdependency token graphs, and strongly connected components to parse logs. Our comprehensive evaluation across various benchmarks demonstrates that Tipping consistently outperforms existing state-of-the-art parsers in both parsing accuracy and processing speed. In the LogPai benchmark, Tipping achieved high group accuracy and $F_1$ scores across diverse datasets, indicating its robustness and versatility. In the LogPM benchmark, Tipping exhibited superior parsing accuracy, parameter mask agreement, and impressive parsing speeds, highlighting its efficiency and precision.

Through experiments conducted using the LogLead tool, Tipping showcased its exceptional ability to handle large datasets efficiently, demonstrating robust parallelization capabilities. It emerged as the fastest parser in the majority of scenarios, particularly excelling in environments with additional CPU cores. This computational efficiency underscores Tipping's scalability, making it a highly viable solution for large-scale log parsing tasks.

Moreover, Tipping's parsing output enhances the downstream task of anomaly detection. Although the differences in anomaly detection performance among various parsers were dataset-dependent, Tipping generally performed well, especially when integrated with supervised learning models. This consistency and adaptability underline Tipping's effectiveness as a log-parsing solution. By offering improvements over existing methods and adaptability to different environments, Tipping proves to be a valuable tool for modern log analysis and anomaly detection tasks.

\section{Benchmark and Tool availability}
For the sake of transparency and accountability, we open-source all repositories and packages used in this paper, including Tipping Rust\footnote{\url{https://github.com/shshemi/tipping-rs}}, Tipping Python\footnote{\url{https://github.com/shshemi/tipping}},
Tipping Pip \footnote{\url{https://pypi.org/project/tipping/}}
LogPai\footnote{\url{https://github.com/M3SOulu/logparser}}, LogPM\footnote{\url{https://github.com/M3SOulu/LogPMBenchmark}}, and LogLead\footnote{\url{https://github.com/EvoTestOps/LogLead}} implementations.

\section{Acknowledgements}
The authors have been supported by the Research Council of Finland (grant IDs: 328058, 349487, 359861)

\bibliographystyle{ieeetr}
\bibliography{refs}

%%%%%%%%%%%%%%%%%%%%%%%%%%%% Biography %%%%%%%%%%%%%%%%%%%%%%%%%%%%%%%%%%%%%%%%%%%%%%
% \newpage

 

% \end{IEEEbiography}



% \vfill

\end{document}